\documentclass[fleqn,usenatbib,useAMS]{mnras}

\usepackage{graphicx}	
\usepackage{amsmath}	
\usepackage{amssymb}	
\usepackage{bm}		
\usepackage{pdflscape}	

\newcommand{\kms}{\,km\,s$^{-1}$} 
\newcommand{\gad}{{\sc Gadget-3}}
\newcommand{\gizmo}{{\sc Gizmo}}

\newcommand{\simba}{{\sc Simba}}
\newcommand{\caesar}{{\sc Caesar}}

\newcommand{\hmpc}{\,h^{-1}{\rm Mpc}}
\newcommand{\HI}{\ion{H}{i}}
\newcommand{\mhi}{M_{\rm HI}}
\newcommand{\fhi}{f_{\rm HI}}

\usepackage[T1]{fontenc}
\usepackage{ae,aecompl}

\usepackage{graphicx}
\usepackage{epstopdf}
\usepackage{subcaption}
\usepackage{caption}
\usepackage{gensymb}
\usepackage{calc}
\usepackage{pdflscape}
\usepackage{rotating}
\usepackage{afterpage}

\newcommand{\scipy}{{\sc scipy}}

\title[]{The baryonic Tully-Fisher relation in the \simba\ simulation}

\author[]{M.~Glowacki$^{1,2}${\thanks{Contact e-mail: \href{mailto:marcin@idia.ac.za}{marcin@idia.ac.za}}, E.~Elson$^{2}$, R.~Dav\'e$^{3,2,4}$}\\
$^{1}$Inter-University Institute for Data Intensive Astronomy, Bellville 7535, South Africa\\
$^{2}$Department of Physics and Astronomy, University of the Western Cape, Robert Sobukwe Road, Bellville 7535, South Africa\\
$^{3}$Institute for Astronomy, Royal Observatory, Univ. of Edinburgh, Edinburgh EH9 3HJ, UK\\
$^{4}$South African Astronomical Observatories, Observatory, Cape Town 7925, South Africa
}
\date{}

\pubyear{2019}

\hypersetup{draft}

\begin{document}
\label{firstpage}
\pagerange{\pageref{firstpage}--\pageref{lastpage}}
\maketitle

\begin{abstract}
We investigate the Baryonic Tully-Fisher Relation (BTFR) in the $(100\hmpc)^3$ \simba\ hydrodynamical galaxy formation simulation together with a higher-resolution $(25\hmpc)^3$ \simba\ run, for over $10,000$ disk-dominated, \HI-rich galaxies.  We generate simulated galaxy rotation curves from the mass distribution, which we show yields similar results to using the gas rotational velocities.  From this we measure the galaxy rotation velocity $V_{\rm circ}$ using four metrics: $V_{\rm max}, V_{\rm flat}, V_{2R_e},$ and $V_{\rm polyex}$. We compare the predicted BTFR to the SPARC observational sample and find broad agreement.  In detail, however, \simba\ is biased towards higher $V_{\rm circ}$ by up to 0.1~dex. We find evidence for the flattening of the BTFR in $V_{\rm circ}>300$~\kms galaxies, in agreement with recent observational findings.  \simba's rotation curves are more peaked for lower mass galaxies, in contrast with observations, suggesting overly bulge-dominated dwarf galaxies in our sample. We investigate for residuals around the BTFR versus \HI\ mass, stellar mass, gas fraction, and specific star formation rate, which provide testable predictions for upcoming BTFR surveys. \simba's BTFR shows sub-optimal resolution mass convergence, with the higher-resolution run lowering $V$ in better agreement with data.
\end{abstract}





\section{Introduction}

In the canonical scenario for disk galaxy formation, cool gas from the halo collapses while approximately conserving specific angular momentum into a rotationally-supported disk~\citep[e.g.][]{FallEfstathiou1980,MoMaoWhite1998}. The kinematics of the resulting disk reflect the kinematics, and thus the mass and radius, of the parent dark matter halo. Hence disk galaxy rotation velocities have long been regarded as a means to connect galaxies with their parent dark matter halos.

Early gas dynamical simulations of galaxy formation, however, did not conform to this simple picture.  Instead, such simulations tended to produce dense stellar systems at early epochs, owing to rapid collapse and hierarchical merging that decoupled the baryonic angular momentum from that of the dark matter.  As a result, these models produced galaxies that had too many stars~\citep[`overcooling';][]{WhiteFrenk1991,Dave2001}, and overly large bulges. Stated another way, it did not seem possible to reconcile hierarchical assembly of galaxies as expected in cold dark matter cosmologies with the conservation of specific angular momentum that was required to match disk galaxy properties.  This became widely known as the `angular momentum catastrophe'~\citep[e.g.][]{NavarroSteinmetz2000,Abadi2003}.  Predicted rotation curves were found to be rapidly rising with strong peaks, rather than slowly rising towards a plateau at larger radii as observed.

While various numerical and dynamical effects were considered towards solving the angular momentum catastrophe, the most broadly accepted solution came from considering the impact of star formation feedback processes, something that was neglected in early simulations.  Since the energy deposition from feedback necessarily occurs where stars form, feedback was able to self-regulate galactic stellar growth in order to produce more realistic stellar contents of galaxies.  Moreover, supernova-generated outflows preferentially carried off gas from galactic centers, thereby lowering bulge fractions and ameliorating the rapid central rise in rotation curves~\citep{Brook2011,Christensen2016}. Since outflows can escape more easily from small potential wells, dwarf galaxies were expected to be most impacted, which thereby suppressed overcooling in early dwarfs, even sometimes yielding bulgeless galaxies~\citep{Governato2007}.  Disk galaxy formation therefore involves a complicated balance between angular momentum loss by violent relaxation versus removal of low-angular momentum gas via outflows, rather that simply halo angular momentum conservation.

The rotation speeds of galaxies are thus an important constraint on galaxy assembly history and feedback processes~\cite[e.g.][]{Haynes1999,Sanders2002,Springob2007,deRossi2010,Ponomareva2018}.  The canonical representation of this is known as the Tully-Fisher relation (TFR), which is a correlation between the luminosity and the rotational velocity of spiral galaxies \citep{Tully1977}.  By including both gas and stars, \citet{McGaugh2012} found that the so-called baryonic TFR (BTFR) provides an even tighter relation spanning many orders of magnitude, suggesting a deep connection between the baryonic content of galaxies and their dark matter halos.

Observing the BTFR typically involves combining optical data to quantify the stellar component, and radio data utilising the atomic hydrogen (\HI\,) spin-flip transition at 21\,cm to probe the gas content \cite[e.g.][]{Verheijen2001,Noordermeer2007,Gurovich2010,McGaugh2012,Zaritsky2014,Lelli2016,Ponomareva2017}. Despite impressive advancements recently in minimising the scatter in the BTFR  \cite[e.g.][]{Lelli2016} such as using mid-infrared emission to probe the total stellar mass content of galaxies, various uncertainties remain, such as an assumed stellar mass-to-light ratio $\Upsilon_{*}$~\citep{McGaugh2012}, and an assumption that the \HI\ rotation speed traces the total mass distribution for the galaxy. Furthermore, it is difficult to detect \HI\ at higher redshifts, with efforts beyond $z$~$>$~0.1 mostly limited to stacking. 

It has also been shown from both real and simulated galaxy samples that the definition of the rotational velocity from rotation curves significantly alters the resulting BTFR \citep{Brook2016,Ponomareva2017,Lelli2019}. Different definitions have been used across observational studies, often owing to differences in the data available, resulting in BTFRs with different slopes and intercepts, and associated intrinsic scatters. This complicates the ability to compare between observations and simulations of galaxy formation, and must be accounted for.

Galaxy formation simulations have long used the TFR and its variants as constraints. Simulations including feedback have broadly been successful at reproducing galaxies that lie on the TFR and BTFR~\citep{Okamoto2005,Governato2007,Guedes2011,Aumer2013,Marinacci2014}. Initially, the importance of numerical resolution was emphasised by e.g. \citet{Governato2007}, both because it provided a stronger coupling of feedback energy to drive outflows, and because it mitigated the artificial viscous angular momentum loss in shear flows endemic to Smooth Particle Hydrodynamics (SPH) schemes.  However, recent models that employ different methods for driving outflows and use more modern hydrodynamics models are able to reproduce the TFR even at $\sim$kpc resolution, such as \citet{Marinacci2014} who used a precursor to the {\sc Illustris} galaxy formation model, or \cite{Ferrero2017} who used the {\sc EAGLE} simulation.  The advantage of using cosmological simulations rather than individual galaxy simulations is in their much larger statistics, and the ability to study correlations with other galaxy properties.  Additionally, galaxy formation simulations can quantify inaccuracies introduced by the aforementioned observational assumptions, and trivially examine redshift evolution.  Modern cosmological simulations thus provide a platform with which to statistically connect the observed BTFR with halo properties.

In this paper we use the \simba\ \citep{Dave2019} suite of cosmological hydrodynamical simulations to extract rotation curves and study the BTFR, for various common definitions of galaxy circular velocity.  We compare our inferred BTFR with observations from the {\em Spitzer} Photometry and Accurate Rotation Curves Survey \cite[SPARCS;][]{Lelli2016}.  We examine the BTFR in various subsamples of our galaxy population, to better understand how rotation speeds vary with galaxy properties, and study how the different parameterisations of the rotation speed affect the BTFR. These results broadly show that \simba\ succeeds at reproducing the rotational properties of galaxies, albeit with notable exceptions.  This establishes \simba\ as a viable platform for exploring and interpreting future surveys  such as Looking At the Distant Universe with the MeerKAT Array \cite[LADUMA;][]{Holwerda2012}, which will investigate the BTFR and how it evolves as a function of redshift.  

This paper is organised as follows.  In Section~\ref{sec:method} we describe \simba\ and outline our sample, our method of generating rotation curves (Section~\ref{sec:rotcurves}), and the rotational velocity definitions examined (Section~\ref{sec:veldefs}). In Section~\ref{sec:results} we compare to SPARCS data, investigate the individual BTFRs for each rotational velocity definition, and explore the dependence of the BTFR on the stellar mass, \HI\ gas fraction, and other properties.  We summarise our findings in Section~\ref{sec:summary}.

\section{Simulations and Analysis}\label{sec:method}
\subsection{\simba}

We employ the \simba\ simulation suite for this analysis~\citep[see][for full description]{Dave2019}.  \simba\ is a cosmological hydrodynamic simulation evolved using the \gizmo\ code~\citep{Hopkins2015}, which itself is an offshoot of \gad~\citep{Springel2005}.  \gizmo\ uses a meshless finite mass (MFM) hydrodynamics solver that is shown to have advantageous features over Smoothed Particle Hydrodynamics and Cartesian mesh codes, such as the ability to evolve equilibrium disks for many dynamical times without numerical fragmentation~\citep{Hopkins2015}, which is desirable for studying the rotation properties of galaxies.

The primary \simba\ simulation is evolved in a $(100\hmpc)^3$ periodic volume, with $1024^3$ dark matter particles and $1024^3$ gas elements.  The assumed cosmology is concordant with~\citet{Planck2016}: \(\Omega_M = 0.3\), \(\Omega_{\Lambda} = 0.7\), \(\Omega_{b} = 0.048\), \(H_0 = 68\) km s\(^{-1}\) Mpch\(^{-1}\), \(\sigma_8 = 0.82\), \(n_s = 0.97\).  This yields a mass resolution of \(9.6\times 10^7 M_{\odot}\) for dark matter particles and \(1.82\times 10^7 M_{\odot}\) for gas elements.  Adaptive gravitational softening length is employed with a minimum \(\epsilon_{\rm{min}} = 0.5h^{-1}\)c\,kpc.

Given the modest spatial resolution of \simba, numerical convergence is an important concern.  To address this, we also analyse a high-resolution \simba\ run having a box size of 25~Mpc~h$^{-1}$, with 512$^3$ dark matter particles and an equal number of gas elements (`\simba-hires'). This run is performed with the exact same input physics, but has 8$\times$ better mass resolution and 2$\times$ better spatial resolution as the fiducial 100~Mpc~h$^{-1}$ run (henceforth \simba-100), albeit in a volume that is 64$\times$ smaller.

\simba\ employs novel state of the art models for sub-resolution processes such as star formation and feedback.  Star formation occurs in molecular gas, with an $H_2$ fraction computed via the prescription of \citet{Krumholz2011} based on metallicity and local column density. The star formation rate (SFR) is calculated from the molecular gas density $\rho_{H_2}$ and the dynamical time \(t_{\textrm{dyn}}\) via \(\textrm{SFR}=\epsilon_* \rho_{H_2}/t_{\textrm{dyn}} \), where \(\epsilon_* = 0.02\) \citep{Kennicutt1998}. The \HI\ fraction of gas particles is computed self-consistently within the code, accounting for self-shielding on the fly based on the prescription in \citet{Rahmati2013}, with a meta-galactic ionizing flux strength assuming a spatially uniform ionising background as specified by \citet{Haardt2012}.  This gives the total shielded gas, and subtracting off the molecular component yields the \HI. Chemical enrichment is tracked for 11 elements, and radiative cooling including metal lines is done via the \textsc{Grackle}-3.1 package~\citep{Smith2017}.

Star formation feedback is included via kinetic decoupled outflows, with kicks applied to gas particles which are then evolved without hydrodynamics until they escape the ISM~\citep{Springel2003}.  The mass loading factor and wind speed follow relations taken from high-resolution Feedback in Realistic Environments (FIRE) zoom simulations~\citep{Muratov2015,Angles2017b}.  Dust formation and destruction is tracked during the simulation evolution, where a fraction of each gas element's metals is locked into dust.

Black hole growth and associated active galactic nuclei (AGN) feedback is included in a unique way in \simba.  Black hole growth is modeled via a torque-limited accretion model~\citep{Hopkins2011,Angles2013,Angles2017} from $T<10^5$K gas since it is intended to model accretion owing to instabilities in a cold gaseous disk, and \citet{Bondi1952} accretion from $T>10^5$\,K gas.  Feedback is purely kinetic and bipolar, in two modes based on the Eddington ratio: outflow speeds of $\sim~1000$\kms at high Eddington ratios (`radiative mode'), increasing to $\sim~10^4$\kms at Eddington ratios below 2\% (`jet mode').  Also, X-ray feedback is included based on the model of \citet{Choi2012}.  Growth occurs primarily via torque-limited accretion, and jet mode is primarily responsible for quenching galaxies.

Galaxies are identified via a 6-D friends of friends (FOF) algorithm. Halos are identified via a 3-D FOF algorithm, but we will not consider halo properties in this work.  Galaxies and halos are cross-matched and their properties computed using \caesar\footnote{\tt https://caesar.readthedocs.io/en/latest/}, a particle-based extension to {\sc yt}~\footnote{\tt https://yt-project.org/}. Galaxy photometry is done using {\sc PyLoser}\footnote{\tt https://pyloser.readthedocs.io/en/latest/}, a Python version of the {\sc Loser} package described in \citet{Dave2017b} that computes the extinction to each star individually based on the line-of-sight dust column density through its host galaxy's gas. {\sc PyLoser} assumes a Chabrier IMF, and uses the Flexible Stellar Population Synthesis models \cite[FSPS;][]{Conroy2009,Conroy2010}, through the python-fsps bindings \citep{Mackey2014}. \HI\ is associated with each galaxy by considering all gas within a galaxy's halo, and summing the \HI\ content of all particles that are most gravitationally bound to that galaxy.  Hence the \HI\ can extend significantly beyond the optical (stellar) radius of the galaxy.

\simba\ reproduces a wide range of observations including stellar growth~\citep{Dave2019}, black hole properties~\citep{Thomas2019}, dust properties~\citep{Li2019}, quenched galaxies properties~\citep{Rodriguez2019}, and most relevantly for this work, cold gas properties~\citep{Dave2020}.  This makes \simba\ a plausible platform for investigating the rotational properties of gas and stars as we do here.

\subsection{Galaxy sample selection}

\begin{figure*}
    \centering
    \includegraphics[width=1.\linewidth]{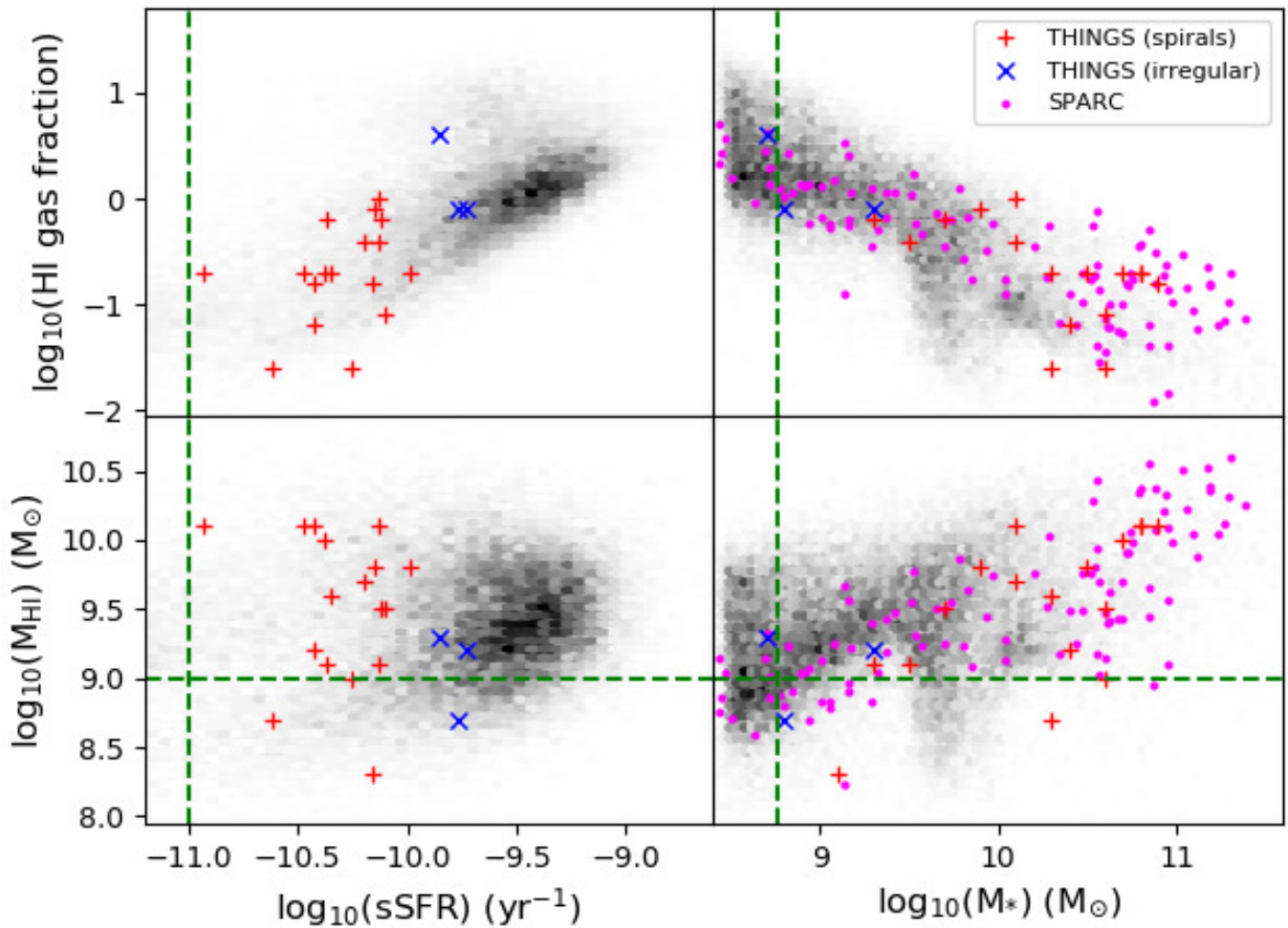}
    \caption{Log-log plots of the comparison of the \HI\ gas fraction, \HI\ mass, stellar mass, and specific star formation rate (sSFR) between \simba\ galaxies within the 100~Mpc~h$^{-1}$ box (black points via hexbin), and THINGS and SPARC galaxies. We have removed three THINGS galaxies with stellar masses below 2\,$\times$\,10$^{8}$\,M$_{\odot}$, as these fall below our stellar mass lower limit for \simba\ galaxies. Blue crosses are irregular morphological type galaxies in THINGS, while red pluses are large spiral galaxies. SPARC galaxies are given as magenta circles in the right side panels. We note that SFR information was not available for the SPARC sample. Through this comparison we constructed a sample selection to select gas-rich large spiral galaxies from \simba\ .}
    \label{fig:Selectioncriteria}
\end{figure*}


Observational BTFR studies with spatially resolved galaxies are most easily performed with large, gas-rich spiral galaxies. A galaxy's \HI\ disk serves as the gravitational tracer well beyond the extent of the stellar disk. However, these studies are restricted to local (low redshift) samples, owing to resolution limitations. Studies of higher redshift samples rely on spectral emission line widths as an estimator of the rotational velocity. In this work, we aim to select simulated galaxies that match the general properties of studies on resolved spiral galaxies. 

Fig.~\ref{fig:Selectioncriteria} shows our sample selection of \simba\ galaxies in stellar versus gas properties at $z=0$ for a 100~Mpc~h$^{-1}$ snapshot. The $y$-axes show the \HI\ fraction $\mhi/M_*$ and the \HI\ mass $\mhi$, while the $x$-axes show the specific star formation rate SFR$/M_*$ and the stellar mass $M_*$. The black hexbins show the \simba\ galaxies, while the points show a comparison to two observational surveys. The first is The \HI\ Nearby Galaxy Survey \cite[THINGS;][]{Walter2008}, with red pluses showing spirals and blue crosses showing irregulars. The other survey is SPARC, which are represented by purple magenta points. We see broad agreement with SPARC galaxies in the $f_{\rm HI}-M_{*}$ and $M_{\rm HI}-M_{*}$ relations (right-side panels), with slightly lower H{\sc i} masses seen in SPARC galaxies at the low stellar mass end. We highlight that no SFR information was available for this sample, and so we restrict its plotting to the right-hand side panels. The stellar mass-size relation, which can play a significant role in the BTFR \citep{Ferrero2017}, has been investigated in fig. 2 of \cite{Appleby2020} at different redshifts, with broad agreement found between the \simba\--100 galaxies and observations at redshift~z~=~0.

To select \simba\ galaxies for our BTFR investigation, we impose the following limits:
\begin{itemize}
    \item $M_{*}$~$>$~5.8$\times$10$^{8}$\,$M_{\odot}$, which is the galaxy stellar mass resolution limit for \simba\ above which we are confident there are enough particles to construct a realistic rotation curve for a galaxy.
    \item $M_{\rm{\HI}}$~$>$~1$\times$10$^{9}$\,$M_{\odot}$, to ensure we are selecting galaxies with sufficient \HI\ content to compare with observations, and construct a rotation curve from the velocity or position information of gas particles in the simulation.
    \item sSFR~$>$~1$\times$10$^{-11}$\,yr$^{-1}$ to select galaxies with active star formation. 
\end{itemize}


Overall, the \simba\ galaxies cover much of the range of of the THINGS sample. In \simba, small galaxies dominate by number, since the dominant selection criterion is just stellar mass. These galaxies tend to have high gas fractions and sSFR compared to observational samples, which tend to be limited by brightness. Nonetheless, in overlapping regions of each space, \simba\ galaxies populate the same region as THINGS galaxies, showing that \simba\ produces reasonable analogs of observed systems in these key quantities. For the 100~Mpc~h$^{-1}$ simulation, we obtain some 11,000 galaxies following the above cuts.  We further remove a small handful of cases of merging galaxies through visual inspection of the moment-0 maps and corresponding rotation curves. For the `\simba-hires' snapshot box, we reduce the stellar and \HI\ mass limits by a factor of 8 to match the increased mass resolution, and obtain another $\sim$650 galaxies.

\subsection{Rotation curves}\label{sec:rotcurves}

For each galaxy within our sample, we make use of the particle positions and velocities to create our rotation curves, through the \caesar\ and {\sc pyGadgetReader} \citep{Thompson2014} tools. We start with the 3D baryonic centre of mass of the galaxy.  We then calculate the distance of each particle associated with the galaxy's halo from that centre of mass position. Stepping out in radius R$_{\rm{i}}$, using a dynamic step size to ensure a sufficient number of particles are included in each bin, we sum the masses M$_{\rm{i}}$ of all dark matter particles within a sphere, and repeat for the stellar and gas particles.

We calculate the total  `ideal' rotational velocity $V_{\rm{total}}$ through quadrature addition of the velocity for each mass component contribution: 
\begin{equation}
V_{\rm{total}}~=~\sqrt{V_{\rm{gas}}^{2} + V_{\rm{star}}^{2} + V_{\rm{DM}}^{2}},
\label{eqn:vel1}
\end{equation}
where for each particle mass component i, the velocity $V_{\rm{j}}$ is calculated through Kepler's Third Law, i.e.
\begin{equation}
V_{\rm{j}}~=~\sqrt{\frac{GM_{\rm{i}}}{R_{\rm{i}}}},
\label{eqn:vel2}
\end{equation}
where G is the gravitational constant and $R_i$ is the radius of particle $i$.

\begin{figure*}
\centering
\begin{minipage}{\textwidth}
\centering
\small
\begin{subfigure}[b]{0.49\textwidth}
  \includegraphics[width=1.\linewidth]{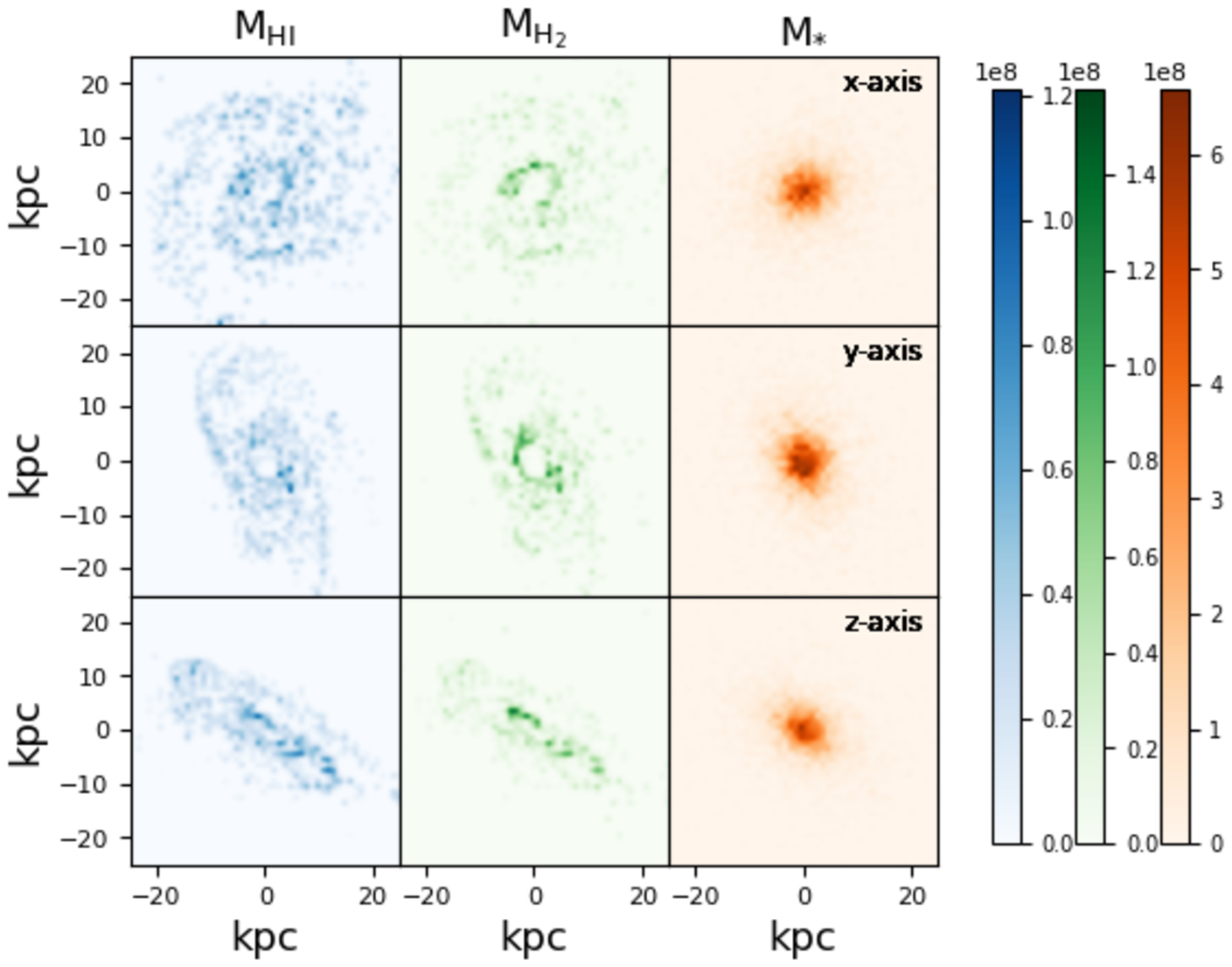}
\end{subfigure}%
\begin{subfigure}[b]{0.49\textwidth}
  \includegraphics[width=1.\linewidth]{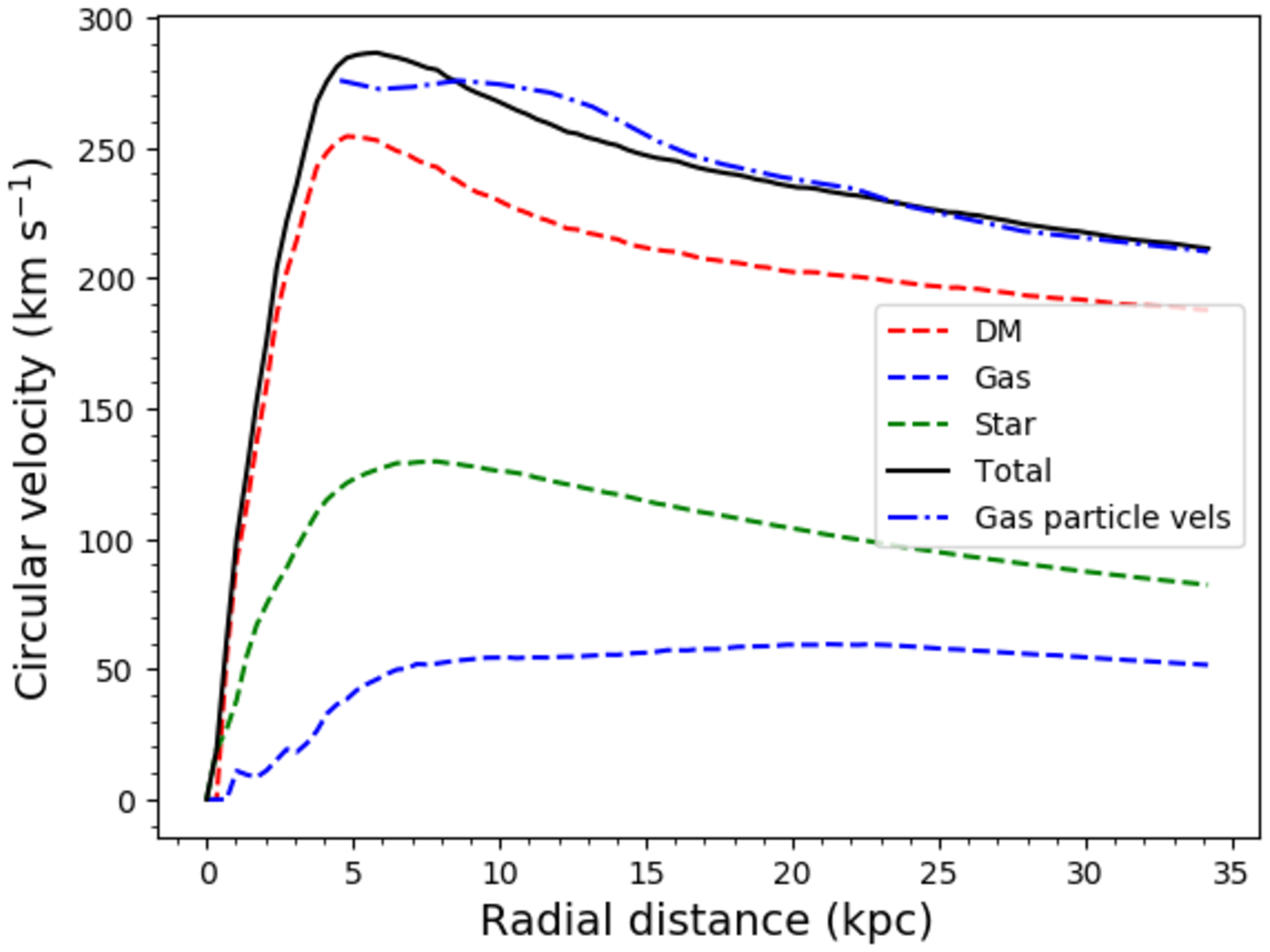}
\end{subfigure}
\begin{subfigure}[b]{0.49\textwidth}
  \includegraphics[width=1.\linewidth]{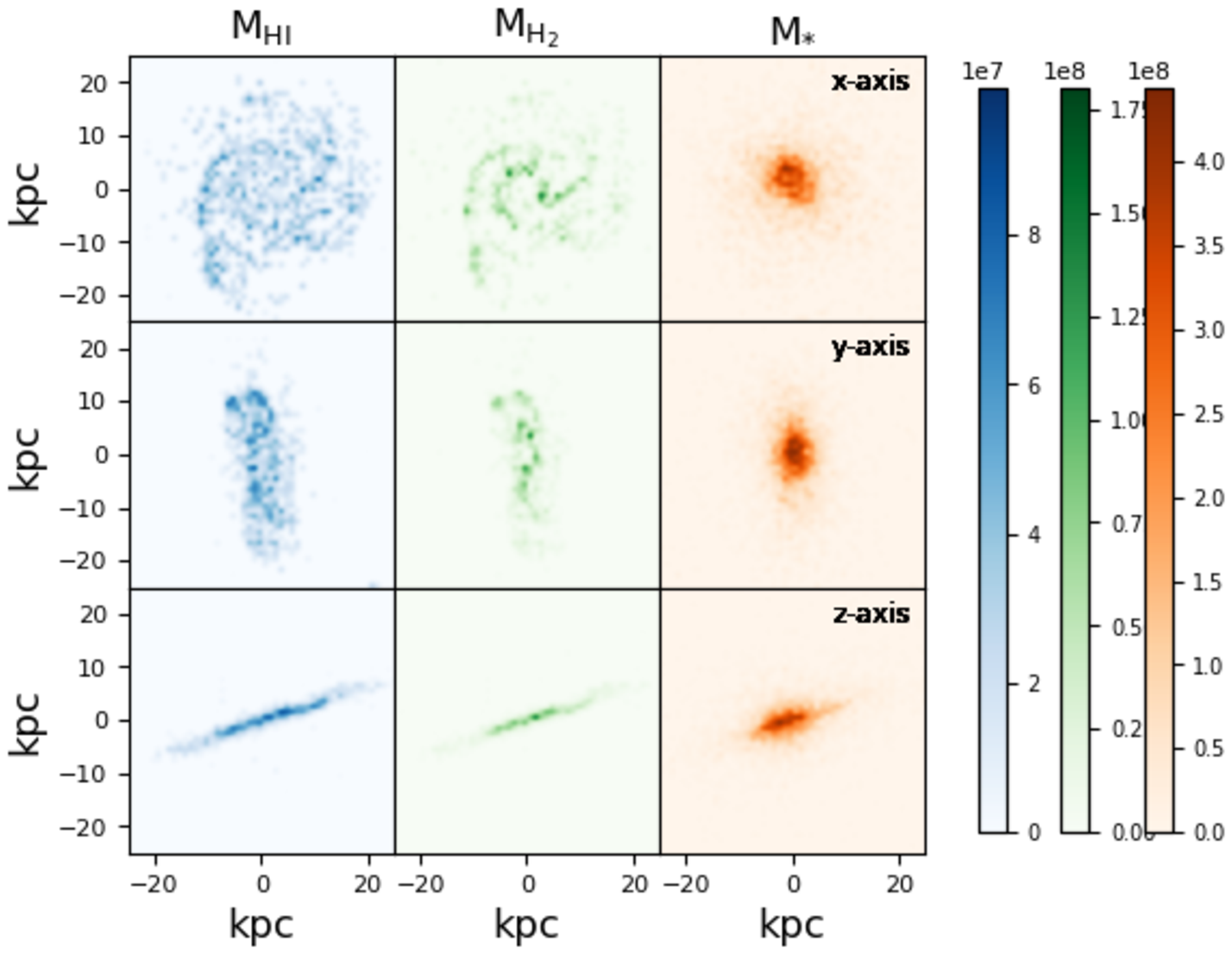}
\end{subfigure}%
\begin{subfigure}[b]{0.49\textwidth}
  \includegraphics[width=1.\linewidth]{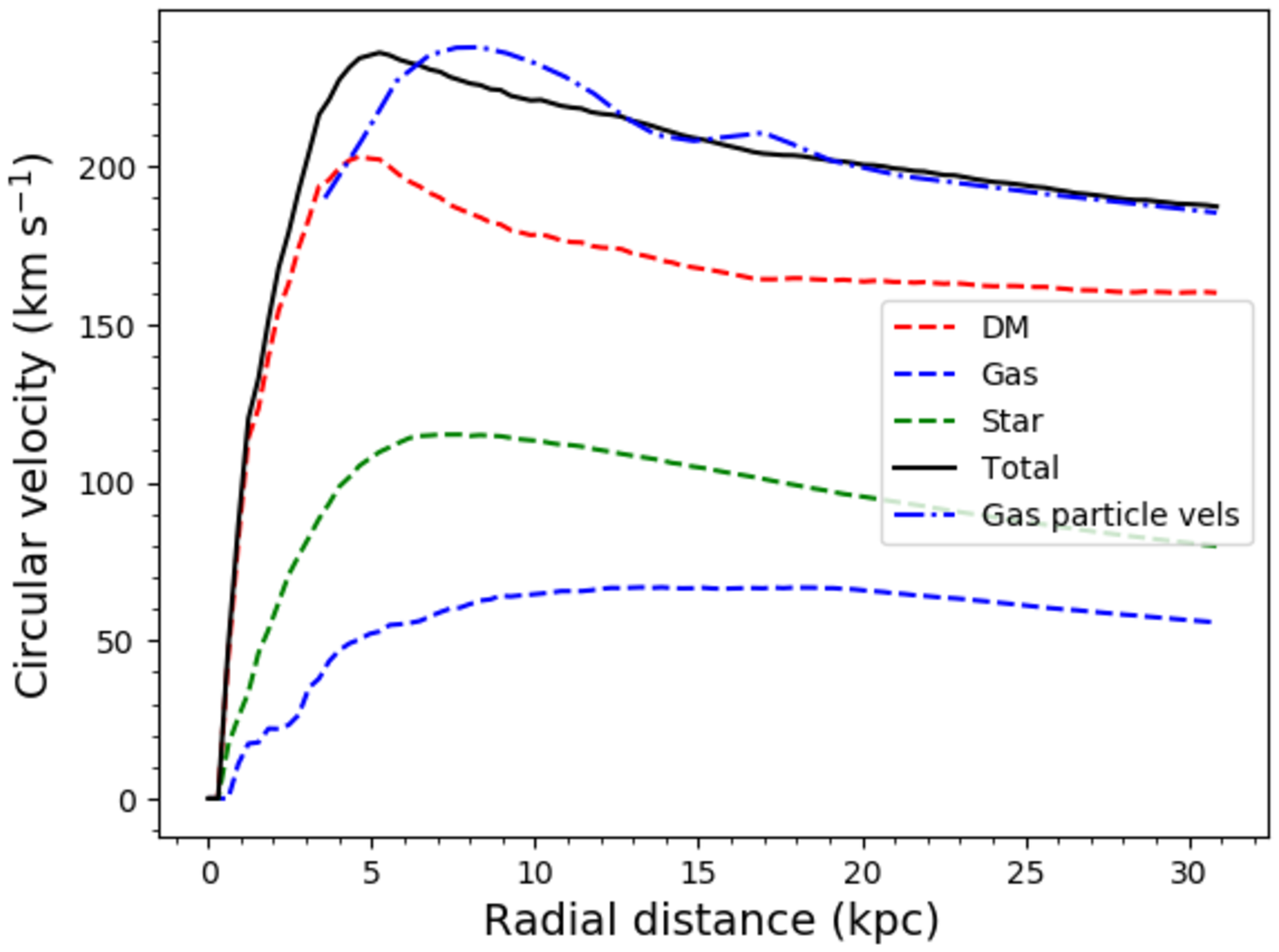}
\end{subfigure}
\begin{subfigure}[b]{0.49\textwidth}
  \includegraphics[width=1.\linewidth]{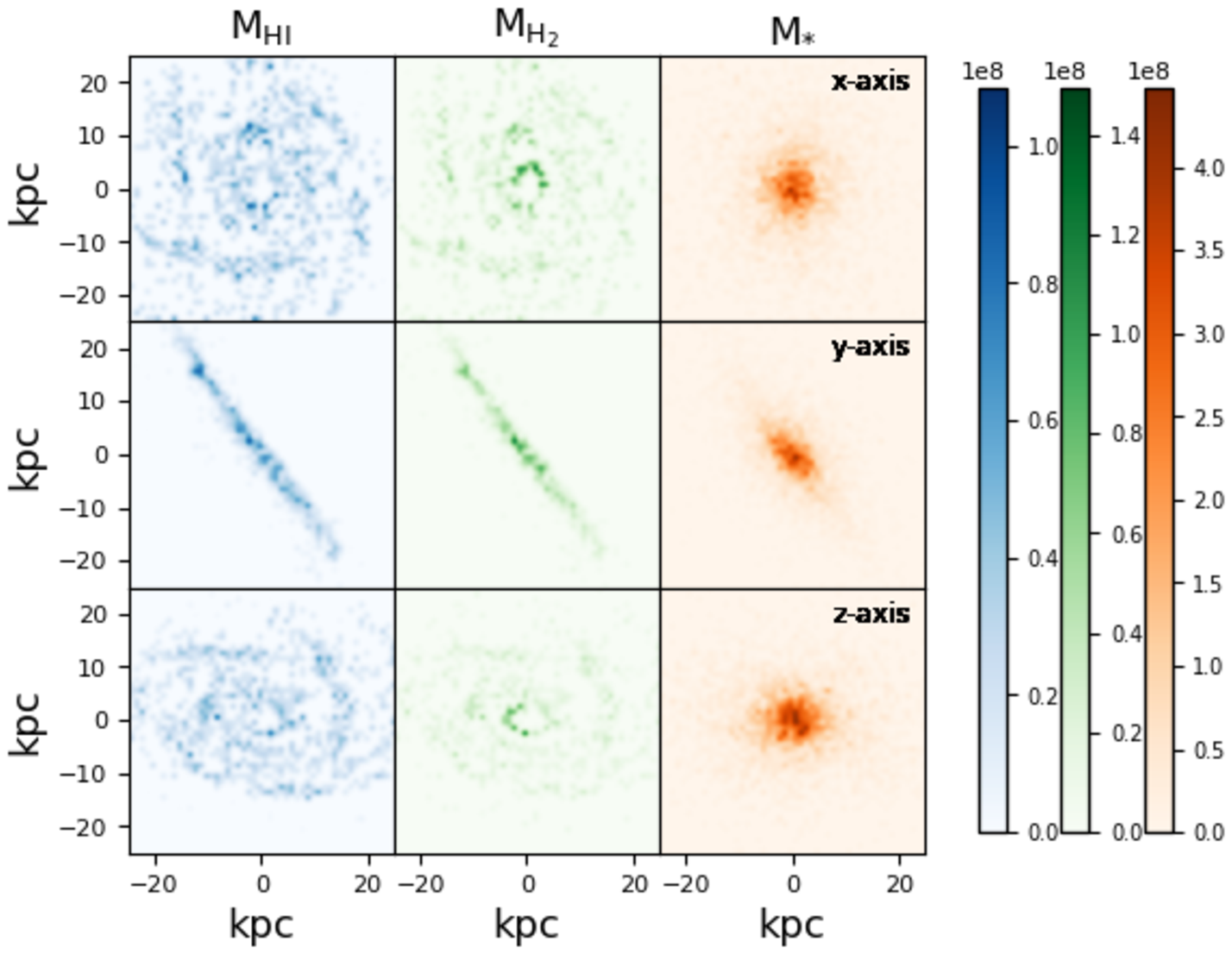}
\end{subfigure}%
\begin{subfigure}[b]{0.49\textwidth}
  \includegraphics[width=1.\linewidth]{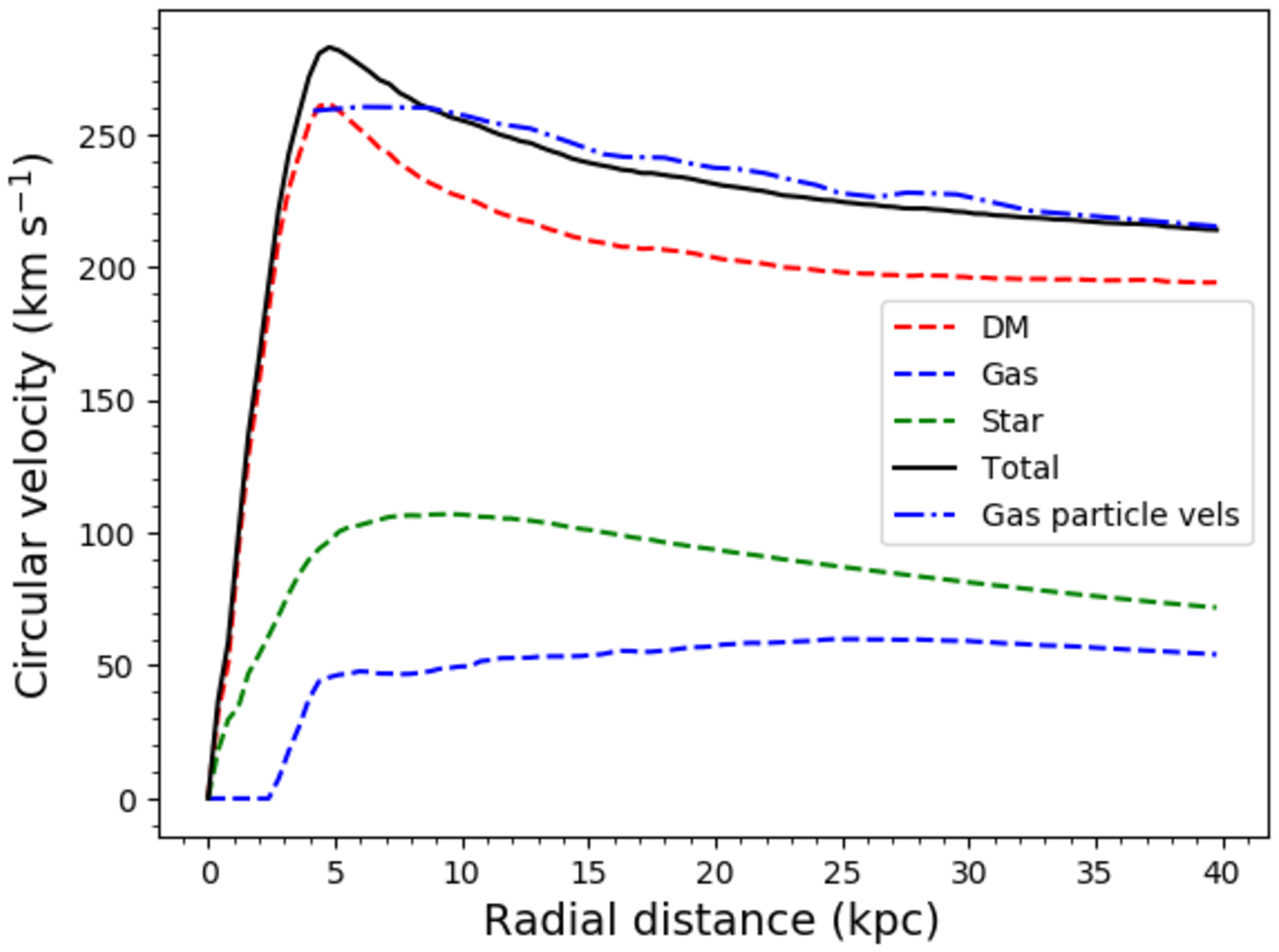}
\end{subfigure}
\caption{Example galaxies from the \simba\ snapshots and their corresponding rotation curves. Galaxies have been selected to closely match galaxies in the THINGS survey, in stellar mass and SFR. Left side: the moment-0 maps for the \HI\,, H$_{2}$, and stellar distributions for the galaxy. Three viewing angles are offered, along the x, y and z-axis of the snapshot box. The colour bars are in units of solar masses M$_{\odot}$. Right side: rotation curves, generated from the mass distribution of all particles making up the galaxy (giving the solid black curve), and alternatively from the velocity of the gas particles (dot-dashed blue curve).}
\label{fig:THINGSexample}
\end{minipage}
\end{figure*}

We test that such `ideal' rotation curves agree with the mean of the total velocities of the gas particles. Within shells of radii stepping out from the centre of mass, for each gas particle we calculate the total velocity within the radius shell, relative to the galaxy's velocity, using
\begin{equation}
V_{\rm{gas~total}}~=~\sqrt{V_{\rm{x}}^{2} + V_{\rm{y}}^{2} + V_{\rm{z}}^{2}},
\label{eqn:velgas}
\end{equation}
where we use the x, y and z components of the gas particle velocity. The mean value of these total velocities is used to construct an alternate rotation curve. Here we adjust the bin size of our radius shells based on the number of gas particles available in the galaxy (e.g. every 100 particles, or 50 in cases of low-mass galaxies), to ensure no low number statistics affect the calculation of the mean total velocity in each radius shell. We see good agreement between the two rotation curve methods. We note this is not a test of whether the H{\sc i} circular velocity is a true measure of the rotation speed; we will examine this in future work using full H{\sc i} data cubes constructed from the \simba\,snapshot files. For the rest of the paper, we use the rotation curves generated through the first method; that is, from the mass distributions of all particles making up each galaxy.

To determine the maximum radius out to which we measure rotational velocities, we use the \HI\ mass-size relation. It has been shown that there exists a tight relation between the \HI\ mass and the extent of the \HI\ disk for spiral galaxies in observational work \citep{Broeils1997,Verheijen2001,Swaters2002,Noordermeer2005,Wang2016}. We apply the relation given in \cite{Wang2016} to the \HI\ masses we have within \simba\ for each galaxy to derive a \HI\ radius.

In Fig.~\ref{fig:THINGSexample}, on the right-hand side we give examples of the two methods of generating rotation curves for a selection of galaxies closely matched to individual THINGS galaxies in $M_*-$SFR space. For these example galaxies, we match to THINGs galaxies with log$_{\rm{10}}$($M_*$) masses of 10.8, 10.6 and 10.7~$M_\odot$, and log$_{\rm{10}}$(SFR) values of
2.123, 3.125, and 2.104~$M_\odot$\,yr$^{-1}$. The `ideal' and velocity-based methods give similar results (solid black and dot-dashed blue curves, respectively). 

We also present the dark matter, stellar, and gas contributions to the `ideal' rotation curve (black) in coloured dashed lines. This shows that the rotation curves are typically dark matter-dominated from quite far in, even for these fairly massive disks.  In other words, the disks are not close to maximal.

The left-hand panels of Fig.~\ref{fig:THINGSexample} the corresponding moment-0 maps (in $M_\odot$kpc$^{-2}$) of the \simba\ galaxies from different viewing angles of the \simba\ snapshot box in \HI\, H$_{\rm{2}}$ and stellar particles.  These were further chosen to be fairly edge-on in one direction, to enable better viewing.  Each shows a thin extended \HI\ and even H$_2$ disk around the stellar portion, which dominates the central region.  These systems also show evidence for a central deficit of cold gas, likely owing to X-ray AGN feedback already present at low levels in these relatively massive systems~\citep{Appleby2020}.

\subsubsection{Stellar mass-to-light ratio}

\begin{figure}
\centering
\begin{subfigure}[b]{0.49\textwidth}
  \includegraphics[width=0.95\linewidth]{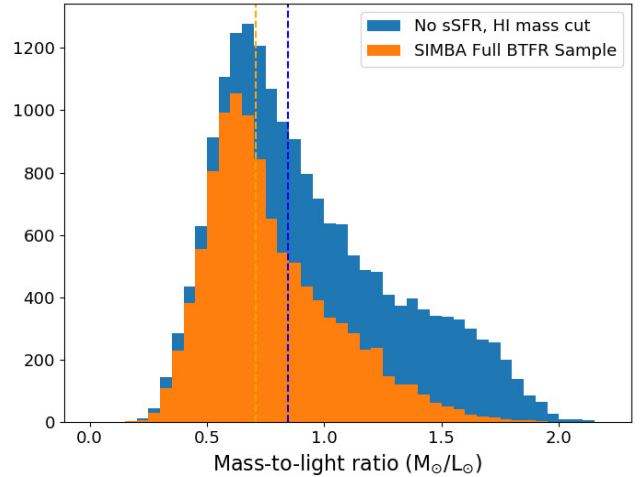}
\end{subfigure}
\caption{The distribution of the stellar mass-to-light ratio for \simba\ galaxies from the 100~Mpc, $z$~=~0 snapshot. We see a peak at $\Upsilon_{*}$~=~0.5, in agreement with the value used by e.g. \citet{Lelli2016}. There is a significant tail extending to higher values.}
\label{fig:masslightratio}
\end{figure}

The stellar mass-to-light ratio ($\Upsilon_{*}$) is the quotient between the total stellar mass  and the luminosity of e.g. a galaxy, and thus it is a necessary quantity to determine the stellar luminosity's contribution to the baryonic mass. This ratio often has to be assumed in observational studies of the BTFR, where the baryonic mass is calculated as the sum of the neutral H{\sc i} gas mass (which is also multiplied by another assumed factor for the molecular gas content), and the product of the stellar luminosity and $\Upsilon_{*}$. \cite{Lelli2016} assume a value of $\Upsilon_{*}$ = 0.5 M$_{\odot}$/L$_{\odot}$ at 3.6~$\micron$ for the SPARC galaxies, based on stellar population synthesis models and colour-magnitude diagrams (see references therein). In \cite{Lelli2017} they also consider separate values for the disk and bulge contribution of galaxies, of $\Upsilon_{\rm{disk}}$ = 0.5~M$_{\odot}$/L$_{\odot}$ and $\Upsilon_{\rm{bulge}}$ = 0.7~M$_{\odot}$/L$_{\odot}$ respectively.

With \simba\ galaxies, we do not need to assume a mass-to-light ratio, as we have the total stellar mass for each simulated galaxy, and can compute the photometry. However, to compare apples-to-apples with the SPARC survey, we include an adjustment factor with our rotation velocity calculation (Eqn.~\ref{eqn:vel1}). We compute the stellar light-to-mass ratio for each \simba\ galaxy via its 3.6~$\mu$m photometry. This is done through {\sc PyLoser}, which computes the apparent and absolute magnitudes of galaxies within cosmological or zoom hydrodynamical galaxy formation simulations, including the option to account for dust extinction to each star particle. We then adjust the contribution of the stellar particles to the total rotation curve to match the assumed 0.5~M$_{\odot}$/L$_{\odot}$ value by \cite{Lelli2016}. 

Fig.~\ref{fig:masslightratio} shows the stellar mass-to-light ratio distribution in for all galaxies with $M_{*}$~$>$~5.8$\times$10$^{8}$\,$M_{\odot}$ in the 100~Mpc~h$^{-1}$ $z$~=~0 \simba\ snapshot in the blue histogram.  The orange histogram shows the sample used to examine the BTFR, with the cuts to sSFR and \HI\ mass described earlier. The distributions peak around 0.6--0.7~M$_{\odot}$/L$_{\odot}$, with a tail to high $\Upsilon_{*}$, and a median $\Upsilon_{*}$ of 0.71 for our BTFR sample. There is a significant proportion of galaxies with higher values than the assumed SPARC $\Upsilon_{*}$ = 0.5~M$_{\odot}$/L$_{\odot}$ value. Hence our adjustment of $\Upsilon_{*}$ is non-trivial when comparing directly with the SPARC survey, affecting the stellar mass contribution by $\approx 40\%$. We use this as motivation to study the BTFR both with and without adjustment to $\Upsilon_{*}$ to see how much this choice can vary the BTFR derived from the simulated galaxies.

However, since the stellar component is usually sub-dominant in the outskirts where we will measure rotation velocities, the overall difference to the BTFR is small.  We show this in Section~\ref{sec:results} and Fig.~\ref{fig:btfr}, where we find only a small difference in the BTFR slope and intercept found across different rotational velocity definitions when applying this mass-to-light ratio adjustment, though it has somewhat more impact on the scatter. We highlight that in fact, fixing our value of $\Upsilon_{*}$ to 0.5 increases the scatter for the BTFR.

Moving forward, in Sections~\ref{sec:btfr_sparc} and \ref{sec:btfr_vel} where we directly compare the \simba\ sample  to the SPARC sample, we show the BTFR where a fixed $\Upsilon_{*}$ = 0.5~M$_{\odot}$/L$_{\odot}$ value is used, although we also give results for the BTFR using the true $\Upsilon_{*}$ value for \simba\ galaxies. Elsewhere, we use the true stellar mass-to-light ratio.

\subsection{Rotational velocities}\label{sec:veldefs}

The BTFR differs both in slope and intercept depending on the particular velocity measure used to parametrise the rotation curve. In order to compare to observational studies, it is thus important to match the definition(s) used. Furthermore, which definition gives the tightest BTFR is a matter of interest, suggesting a closer connection between that measure and halo properties. \cite{Lelli2019} explored this subject with multiple methods for both rotation curves and \HI\ spectral line widths for the SPARC sample, and determined that for their sample, the $V_{\rm{flat}}$ method (i.e. the average velocity of the flat/tail end of the rotation curve) gave the tightest -- and steepest -- relation with a scatter of 0.026$\pm$0.007 dex.

We thus consider four methods for selecting the rotational velocity from the rotation curve based on studies by \cite{Brook2016,Ponomareva2017,Lelli2019}:

\begin{itemize}
\item The circular velocity $V_{\rm{2R_{e}}}$ at twice the effective radius of the galaxy. The effective radius is defined as that enclosing half the stellar mass in the galaxy.

\item The circular velocity $V_{\rm{max}}$, measured at the peak of the ideal rotation curve. 

\item The average circular velocity along the flat part of the rotation curve $V_{\rm{flat}}$, defined in \cite{Lelli2016} as the average of outermost points of a rotation curve with relative cumulative differences smaller than 5\% in rotational velocity.

\item The `Polyex' rotational velocity parametric model $V_{\rm{polyex}}$, as defined from 
\cite{Giovanelli2002} as
\begin{equation}
V(r) = V_{0}(1 - e^{-r/r_{\rm{pe}}})(1 - \alpha r/r_{\rm{pe}}),    
\end{equation}
where $V_{\rm{0}}$ regulates the overall amplitude of the rotation curve, $r_{\rm{pe}}$ is the scale length for the inner steep rise of the rotation curve, and yields a scale length for the inner steep rise, and $\alpha$ sets the slope of the slowly varying outer part. We take the maximum of the Polyex curve as our measure. The \scipy\ \citep{Virtanen2019} curve fitting routine in Python was used to determine the best fitting Polyex curve. 
\end{itemize}

\begin{figure}
\centering
\begin{subfigure}[b]{0.49\textwidth}
  \includegraphics[width=0.95\linewidth]{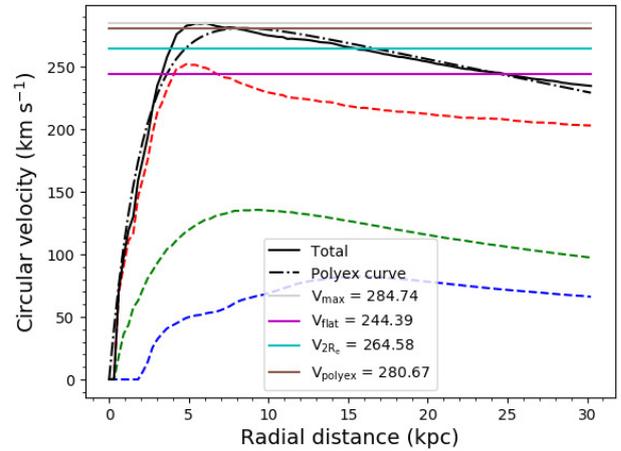}
\end{subfigure}%
\caption{Four different definitions of rotational velocity from an example galaxy rotation curve; $V_{\rm{max}}$ (gray horizontal line), $V_{\rm{flat}}$ (magenta), $V_{\rm{2R_{e}}}$ (cyan), and $V_{\rm{polyex}}$ (brown). The best-fit Polyex function to the total rotation curve (solid black line) is given as a dot-dashed black line for comparison. The chosen methods results in a significantly different value, and hence a different slope and y-intercept, for the corresponding BTFR.}
\label{fig:rotvelcomp}
\end{figure}

Fig.~\ref{fig:rotvelcomp} illustrates these various definitions for a typical disk galaxy rotation curve. The solid black line shows the ideal rotation curve, as calculated as the quadratic sum of the individual mass components (gas, dark matter and stars, in dashed lines). The dot-dashed black line shows the best-fit Polyex curve. The different values for the rotation velocity are indicated by the various horizontal lines, colour-coded as labeled.

$V_{\rm{max}}$ (grey) usually has the highest values, almost by definition. $V_{\rm{polyex}}$ (brown) also has quite high values in general, since it is also fitting a peak, though the curve-fitting procedure typically smooths out the peak slightly, so its values are typically slightly below $V_{\rm{max}}$. $V_{\rm{2R_{e}}}$ is often the next highest, because the rotation curves continue to drop outside of this radius. As such, $V_{\rm{flat}}$ measured in the outskirts generally yields the lowest values.

As in \cite{Lelli2019}, we make linear fits to the BTFR of the form 
\begin{equation}
    \log_{\rm{10}}(M_{\rm{bar}}) = m\,\log_{\rm{10}}\left(\frac{V}{\rm 200\; km\;s^{-1}}\right) + b_{200},
\end{equation}
where $M_{\rm{bar}}$ is the baryonic mass in $M_{\odot}$ for the galaxy calculated to the extent of the rotation curve (i.e. the distance determined by the \HI\ mass-size relation), $V$ is one of the above four velocity definitions in \kms, and $m$ and $b_{200}$ are the slope and intercept normalised at 200\kms.

\begin{figure*}
\centering
\begin{subfigure}[b]{0.99\textwidth}
  \includegraphics[width=1.\linewidth]{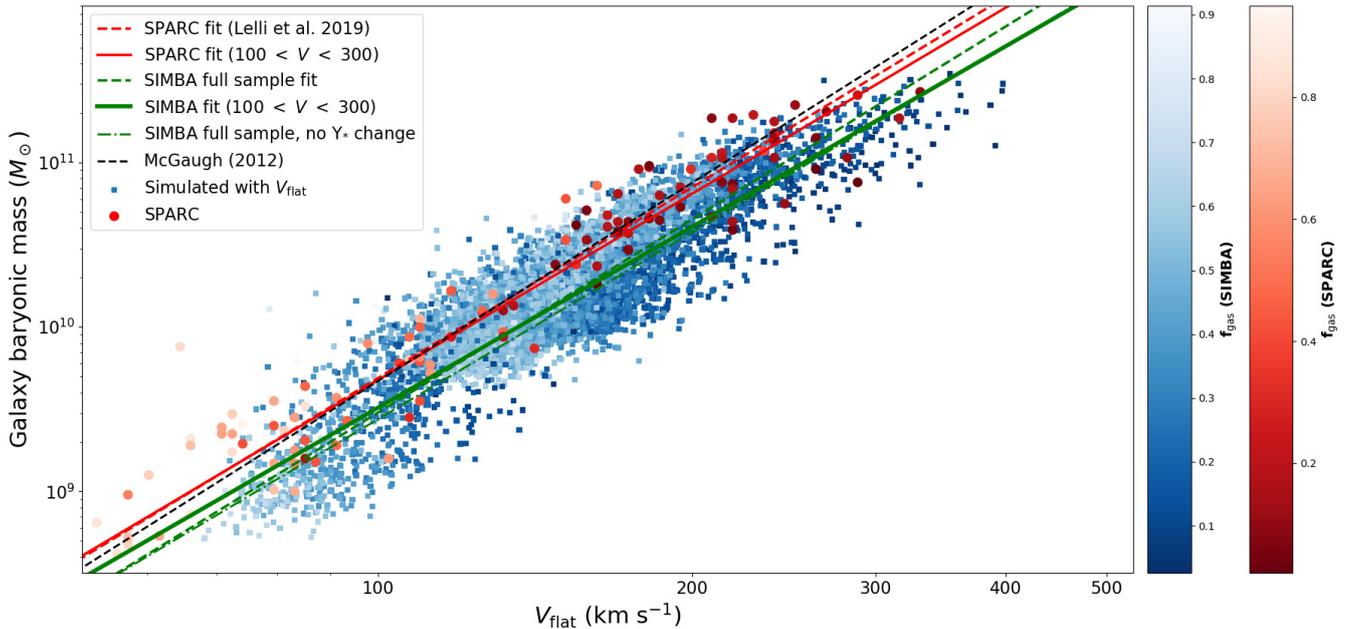}
\end{subfigure}%
\caption{Comparison of the SPARC BTFR (red, large coloured circles), with our simulated galaxies' BTFR (blue, smaller coloured squares) down to stellar masses of 5.8$\times$10$^{8}$\,$M_{\odot}$  for \simba\--100 galaxies, and 8 times lower (7.25$\times$10$^{7}$\,$M_{\odot}$) for the high resolution galaxies. The colour scales give the gas fractions for both samples. To best compare with the SPARC work, we take the $V_{\rm{flat}}$ velocity definition for our sample, as defined in \citet{Lelli2016} and \citet{Lelli2019}. At the higher velocity end (ergo, typically more massive galaxies) we find our sources lie below the trend, indicating a turn-off in the relation for massive galaxies in our simulation. We give the best fits to both our sample (green lines) and the SPARC (red lines) sample for the full velocity range, and between 100~$<$~$V$~$<$~300~km\,s$^{-1}$. We also include the \citet{McGaugh2012} result (black dashed curve), and the best fit (green dot-dashed line) to the full sample without the adjustment to $\Upsilon_{*}$ to compare to the SPARC sample, which results in a similar fit.}
\label{fig:sparccompare}
\end{figure*}

To quantify the scatter, we follow \cite{Lelli2019} and use the orthogonal maximum likelihood (ML) method, which assumes the intrinsic scatter $\sigma_{\perp}$ is Gaussian along the perpendicular direction to the best fitting line.  In considering the full sample where we combine
the 100~Mpc~h$^{-1}$ volume and \simba-hires, we give higher weighting of a factor of 8 to the \simba-hires galaxies, in order to account for the factor of eight higher mass resolution it offers over the \simba\--100 box. This is incorporated by use of the standard affine-invariant ensemble sampler in {\sc emcee} \citep{Foreman-Mackey2013}, which includes inverse variance weighting. Estimation of the error on the scatter and other variables is performed with 100 random walkers. 

\section{Baryonic Tully-Fisher Relation Results}\label{sec:results}

In this section we compare \simba's BTFR with that found in \cite{Lelli2016} and \cite{Lelli2019}, and compare each velocity definition given in Section~\ref{sec:veldefs} with each other in regards to slope, intercept, and scatter.

\subsection{The BTFR in \simba\ versus SPARC}\label{sec:btfr_sparc}

\begin{table*}
    \centering
    \begin{tabular}{lcccc}
    \hline 
Sample and velocity definition & $m$ & $b$ & log$_{\rm{10}}$($b_{200}$) & $\sigma_{\perp}$ \\
\hline
Full sample, 100~$<$~$V$~$<$~300~km\,s$^{-1}$: \\ 
$\sim$10,500 sources\\
\hline
$V_{\rm{2R_{e}}}$    & 3.78$\pm$0.02 & 1.86$\pm$0.05 & 10.58$\pm$0.08 & 0.075$\pm$0.001 \\
$V_{\rm{max}}$       & 4.19$\pm$0.01 & 0.75$\pm$0.02 & 10.40$\pm$0.05 & 0.068$\pm$0.001 \\
$V_{\rm{flat}}$      & 3.65$\pm$0.02 & 2.21$\pm$0.03 & 10.62$\pm$0.07 & 0.070$\pm$0.001 \\
$V_{\rm{polyex}}$    & 3.36$\pm$0.01 & 2.70$\pm$0.04 & 10.44$\pm$0.06 & 0.078$\pm$0.002 \\
\hline
\simba\--100, 140~$<$~$V$~$<$~300~km\,s$^{-1}$: \\ 
$\sim$9000 sources\\
\hline
$V_{\rm{2R_{e}}}$    & 3.40$\pm$0.05 & 2.56$\pm$0.11 & 10.39$\pm$0.25 & 0.074$\pm$0.006 \\
$V_{\rm{max}}$       & 3.51$\pm$0.04 & 2.11$\pm$0.04 & 10.19$\pm$0.17 & 0.063$\pm$0.003 \\
$V_{\rm{flat}}$      & 3.24$\pm$0.05 & 3.02$\pm$0.12 & 10.47$\pm$0.25 & 0.082$\pm$0.003 \\
$V_{\rm{polyex}}$    & 3.76$\pm$0.04 & 1.64$\pm$0.08 & 10.30$\pm$0.17 & 0.056$\pm$0.003 \\
\hline
High resolution sample, 100~$<$~$V$~$<$~300~km\,s$^{-1}$: \\
$\sim$500 sources\\
\hline
$V_{\rm{2R_{e}}}$    & 4.80$\pm$0.11 & -0.29$\pm$0.21 & 10.76$\pm$0.47 & 0.052$\pm$0.005 \\
$V_{\rm{max}}$       & 5.07$\pm$0.08 & -1.11$\pm$0.19 & 10.57$\pm$0.56 & 0.034$\pm$0.004 \\
$V_{\rm{flat}}$      & 4.24$\pm$0.10 & 0.98$\pm$0.21 & 10.72$\pm$0.74 & 0.067$\pm$0.007 \\
$V_{\rm{polyex}}$    & 4.47$\pm$0.11 & 0.38$\pm$0.24 & 10.67$\pm$0.53 & 0.035$\pm$0.004 \\
\hline
Full sample, all velocities: \\ 
$\sim$11,000 sources\\
\hline
$V_{\rm{2R_{e}}}$    & 3.53$\pm$0.01 & 2.39$\pm$0.03 & 10.52$\pm$0.06 & 0.083$\pm$0.001 \\
$V_{\rm{max}}$       & 4.34$\pm$0.01 & 0.44$\pm$0.02 & 10.42$\pm$0.04 & 0.069$\pm$0.001 \\
$V_{\rm{flat}}$      & 3.90$\pm$0.01 & 1.68$\pm$0.02 & 10.65$\pm$0.05 & 0.069$\pm$0.001 \\
$V_{\rm{polyex}}$    & 3.46$\pm$0.01 & 2.49$\pm$0.03 & 10.46$\pm$0.05 & 0.077$\pm$0.001 \\
\hline
Full sample, no $\Upsilon_{*}$ change, all velocities: \\ 
$\sim$11,000 sources\\
\hline
$V_{\rm{2R_{e}}}$    & 3.53$\pm$0.01 & 2.36$\pm$0.03 & 10.48$\pm$0.06 & 0.081$\pm$0.001 \\
$V_{\rm{max}}$       & 4.18$\pm$0.01 & 0.73$\pm$0.02 & 10.36$\pm$0.04 & 0.064$\pm$0.001 \\
$V_{\rm{flat}}$      & 3.77$\pm$0.01 & 1.90$\pm$0.02 & 10.58$\pm$0.04 & 0.068$\pm$0.001 \\
$V_{\rm{polyex}}$    & 3.43$\pm$0.01 & 2.52$\pm$0.03 & 10.41$\pm$0.05 & 0.072$\pm$0.001 \\
\hline
Full sample, $V$~$>$~300~km\,s$^{-1}$: \\ 
$\sim$100 sources\\
\hline
$V_{\rm{2R_{e}}}$    & 2.41$\pm$0.71 & 5.15$\pm$1.78 & 10.69$\pm$0.65 & 0.081$\pm$0.031 \\
$V_{\rm{max}}$       & 2.59$\pm$0.43 & 4.60$\pm$1.08 & 10.56$\pm$0.61 & 0.074$\pm$0.041 \\
$V_{\rm{flat}}$      & 2.60$\pm$0.66 & 4.68$\pm$1.68 & 10.66$\pm$0.60 & 0.071$\pm$0.038 \\
$V_{\rm{polyex}}$    & 3.10$\pm$0.68 & 3.41$\pm$1.76 & 10.53$\pm$0.49 & 0.049$\pm$0.042 \\
\hline
\end{tabular}
    \caption{BTFRs for different rotational velocity definitions, when considering the full (combined) sample, the \simba\--100~Mpc~h$^{-1}$ snapshot box, and the high-resolution box. We give the slope ($m$), y-intercept ($b$), the logarithm of the y-intercept at 200~km\,s$^{-1}$ (log$_{\rm{10}}$($b_{200}$)), and the orthogonal intrinsic scatter ($\sigma_{\perp}$) in dex for log$_{\rm{10}}$(M$_{\rm{bar}}$) = m\,log$_{\rm{10}}$($V$) + b. For rotational velocity cuts, we use either 140~$<$~$V$~$<$~300~km\,s$^{-1}$ or 100~$<$~$V$~$<$~300~km\,s$^{-1}$ limits (removing super-massive galaxies and those with limited stellar masses, ergo less particles making up the galaxy), with the 140~$<$~$V$~$<$~300~km\,s$^{-1}$ cut reserved for the \simba\--100 sample which lacks the lower mass, and hence slower rotating, galaxies offered by the high-resolution sample. We also give the full combined sample with no rotational velocity limits, and with $V$~$>$~300~km\,s$^{-1}$ (just super-massive galaxies). We find that for our sample the $V_{\rm{2R_{e}}}$ velocity definition gives the tightest BTFR when no mass-to-light ratio adjustment is made, but with it to better compare to SPARC $V_{\rm{flat}}$ has the tightest relation. Slopes for each rotational velocity relation get steeper when removing the lower and higher velocity end of the sample (see also Fig.~\ref{fig:btfrstats}). We also observe a flatter BTFR for the most massive galaxies in our sample, which was also seen observationally \citep{Ogle2019}.}
    \label{tab:btfr_fits}
\end{table*}
\begin{table*}
\ContinuedFloat
    \centering
    \begin{tabular}{lcccc}
    \hline 
Sample and velocity definition & $m$ & $b$ & log$_{\rm{10}}$($b_{200}$) & $\sigma_{\perp}$ \\
\hline
Full sample, B/T~$<$~0.3, 100~$<$~$V$~$<$~300~km\,s$^{-1}$: \\ 
$\sim$6,000 sources\\
\hline
$V_{\rm{2R_{e}}}$    & 3.78$\pm$0.02 & 1.92$\pm$0.04 & 10.63$\pm$0.09 & 0.077$\pm$0.001 \\
$V_{\rm{max}}$       & 4.32$\pm$0.01 & 0.51$\pm$0.03 & 10.46$\pm$0.06 & 0.074$\pm$0.001 \\
$V_{\rm{flat}}$      & 3.68$\pm$0.02 & 2.18$\pm$0.04 & 10.65$\pm$0.08 & 0.073$\pm$0.001 \\
$V_{\rm{polyex}}$    & 3.51$\pm$0.02 & 2.45$\pm$0.03 & 10.52$\pm$0.07 & 0.080$\pm$0.001 \\
\hline
Full sample, $\kappa_{\rm rot}$~$>$~0.5, 100~$<$~$V$~$<$~300~km\,s$^{-1}$: \\ 
$\sim$4,700 sources\\
\hline
$V_{\rm{2R_{e}}}$    & 3.54$\pm$0.02 & 2.49$\pm$0.05 & 10.64$\pm$0.10 & 0.082$\pm$0.002 \\
$V_{\rm{max}}$       & 4.13$\pm$0.02 & 0.96$\pm$0.03 & 10.46$\pm$0.07 & 0.079$\pm$0.001 \\
$V_{\rm{flat}}$      & 3.51$\pm$0.02 & 2.59$\pm$0.04 & 10.66$\pm$0.08 & 0.079$\pm$0.001 \\
$V_{\rm{polyex}}$    & 3.40$\pm$0.03 & 2.74$\pm$0.04 & 10.54$\pm$0.08 & 0.084$\pm$0.001 \\
\hline
\end{tabular}
    \caption{Continued. The best-fitting BTFR parameters \simba\ galaxies in the combined sample where simple morphological estimator cuts are made to remove bulge-dominated galaxies.}
\end{table*}

We begin by comparing our BTFR with that of the SPARC survey, using the $V_{\rm{flat}}$ velocity definition. 

Fig.~\ref{fig:sparccompare} shows the BTFR from \simba\ galaxies (in both the 100~Mpc~h$^{-1}$ and high resolution snapshots), versus that from the SPARC survey, using the $V_{\rm{flat}}$ velocity definition that was deemed by \cite{Lelli2019} to provide the tightest BTFR. \simba\ galaxies are shown as squares, coloured in blue by the gas fraction of each galaxy. The SPARC data are shown as circles, coloured in red by their measurement for gas fraction. Lines indicate the best-fit power law to \simba\ galaxies overall (green dashed) and the SPARC sample overall (red dashed), and \simba\ galaxies with 100~$<$~$V_{\rm{flat}}$~$<$~300~km\,s$^{-1}$ (solid green), and the corresponding SPARC fits (solid red). We also give the best fit (green dot-dashed line) to the full sample without the adjustment to $\Upsilon_{*}$, and the fit found in \cite{McGaugh2012}.

The \simba\ galaxy sample is shifted to higher rotation velocities versus SPARC's, so it is important to only conduct comparisons in an overlapping velocity range.  We choose the velocity range of 100~$<$~$V_{\rm{flat}}$~$<$~300~km\,s$^{-1}$ to compare fits because it is the region where the samples overlap. The lower limit is set somewhat above the smallest rotation velocities in \simba\ in order to avoid Eddington bias in the fits. As our sample is limited to $M_*>5.8\times 10^{8} M_{\odot}$, this roughly translates into a lower limit in baryonic mass of $\sim 4\times 10^9 M_\odot$ given the high gas fractions in low-mass galaxies~\citep{Dave2020}. While \simba\ produces galaxies down to $V\sim 80$\kms, the scatter around the relation means that including such galaxies would bias the intercept towards higher values. The upper limit is set as roughly the upper limit of the SPARC sample. We do however give the fit to the full sample with no velocity cut as well (dashed green line), and find good agreement there, albeit with an offset in the intercept between our sample and that of SPARC. Between the two \simba\ fits given here, the main difference is in the slope, while we find similar $b_{\rm{200}}$ values.

It has been observed that `super spiral' galaxies with rotational velocities above 300~km\,s$^{-1}$ result in a significantly flatter BTFR \citep{Ogle2019}. They found that super spirals with velocities greater than 340~km\,s$^{-1}$ are undermassive for their dark matter halos, and found a best BTFR fit slope of 1.64~$\pm$~0.30. Indeed, in the next section we shall see that \simba\ yields a qualitatively similar result, albeit at low significance. Hence by excluding the lower and higher velocity end of our sample, we obtain a steeper BTFR slope than that found when including these super spiral galaxies.

Comparing \simba\ to the SPARC sample within the 100--300~km\,s$^{-1}$ velocity range, we find reasonable agreement. The \simba\ slope is $3.53\pm 0.02$, while the SPARC slope given in \cite{Lelli2019} is $3.85\pm 0.09$, where it was noted that `systematic uncertainties may drive the slope from 3.5 to 4.0'. There is a mild tension in that \simba\ predicts a somewhat shallower slope, and also a lower intercept at 200\kms. Overall, however, the agreement in both slope and intercept when utilising the same method for computing $V_{\rm flat}$ is a solid success for \simba, and shows that the connection between rotation speed in the outskirts and baryonic mass is well reproduced in this model.

\subsection{BTFR dependence on the velocity definition}\label{sec:btfr_vel}

\begin{figure*}
\centering
\begin{subfigure}[b]{0.99\textwidth}
  \includegraphics[width=1.\linewidth]{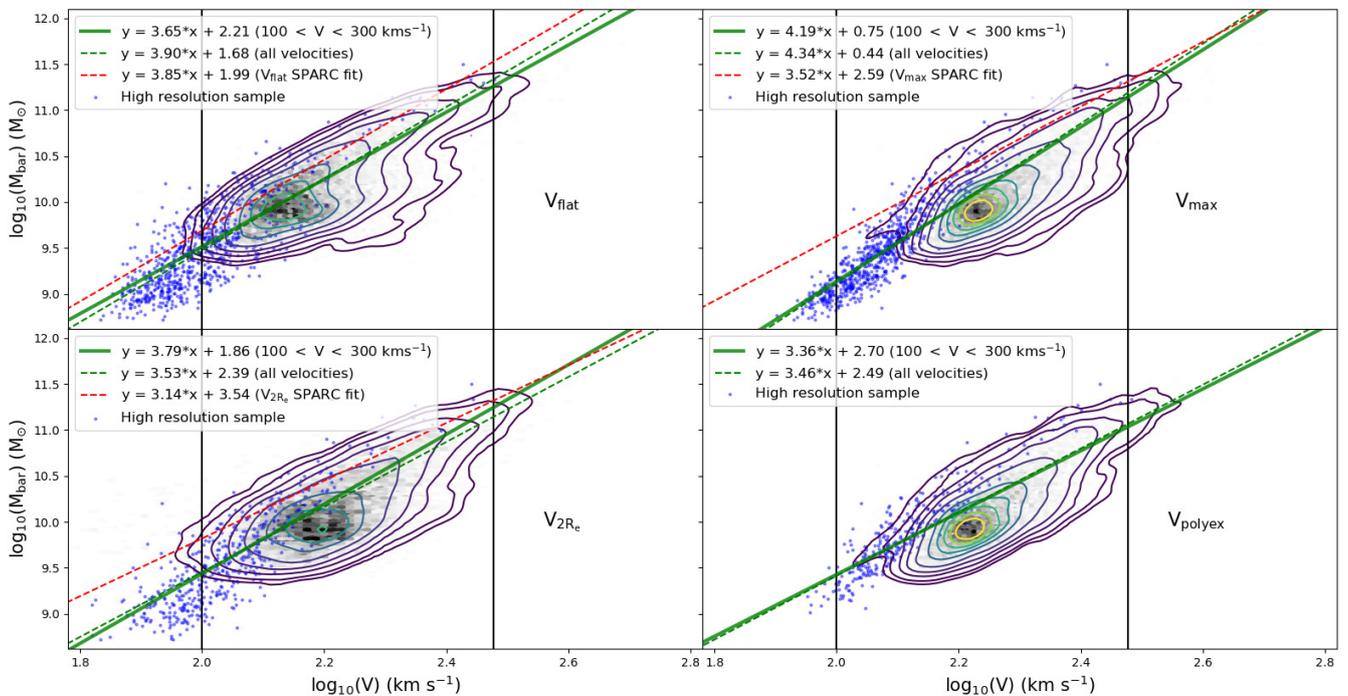}
\end{subfigure}
\caption{The BTFR for our galaxies, using different rotational velocity definitions and with an $\Upsilon_{*}$ correction applied to match the SPARC study. The black hexbins show the individual galaxies in the full (combined) sample, with contours enclosing 0.1, 0.2, 0.5, 1, 2, 4, 6, 8, 10, and 12-$\sigma$ of the Gaussian kernel density estimation of the \simba-100 galaxy distribution. Blue dots indicate the galaxies from the high-resolution subsample. Our fit (red solid line) is calculated from the sample within a velocity cut indicated by the black vertical lines (between 100 and 300 km\,s$^{-1}$), while the dashed red line has no velocity cut. Inclusion of the higher mass galaxies, ergo the faster rotating ones, slightly decreases the slope. We compare with the fits found in \citet{Lelli2019} (green dashed line) for $V_{\rm{flat}}$, $V_{\rm{max}}$, and $V_{\rm{2R_{e}}}$. }
\label{fig:btfr}
\end{figure*}

\begin{figure*}
\centering
\begin{subfigure}[b]{0.99\textwidth}
  \includegraphics[width=1.\linewidth]{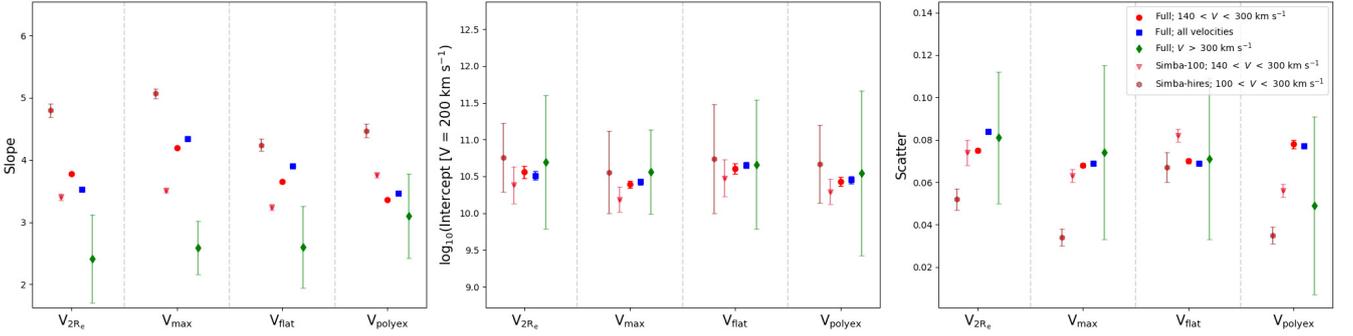}
\end{subfigure}%
\caption{The BTFR best-fit values for the different velocity definitions we examined in this study, with the mass-to-light ratio adjustment to match the SPARC survey. The left panel gives the best-fit slope, the middle panel the logarithm of the best-fit intercept at $V$~=~200~km\,s$^{-1}$, and the right panel the orthogonal scatter. Points are color-coded by the velocity cuts used in selecting the sample, if any, and the sample considered. The slope of the BTFR decreases in steepness for the $V$~$>$~300~km\,s$^{-1}$ massive galaxy subsample (green diamond), relative to the 100~$<$~$V$~$<$~300~km\,s$^{-1}$ subsample (red circle) and the full sample (blue square). We see larger errors due to the far smaller sample for the massive galaxy subsample. We also compare similar velocity cuts for the full (combined) sample with the individual \simba\--100 and high-resolution samples (red triangles and hexagons respectively).}
\label{fig:btfrstats}
\end{figure*}

\begin{figure}
    \centering
  \includegraphics[width=0.46\textwidth]{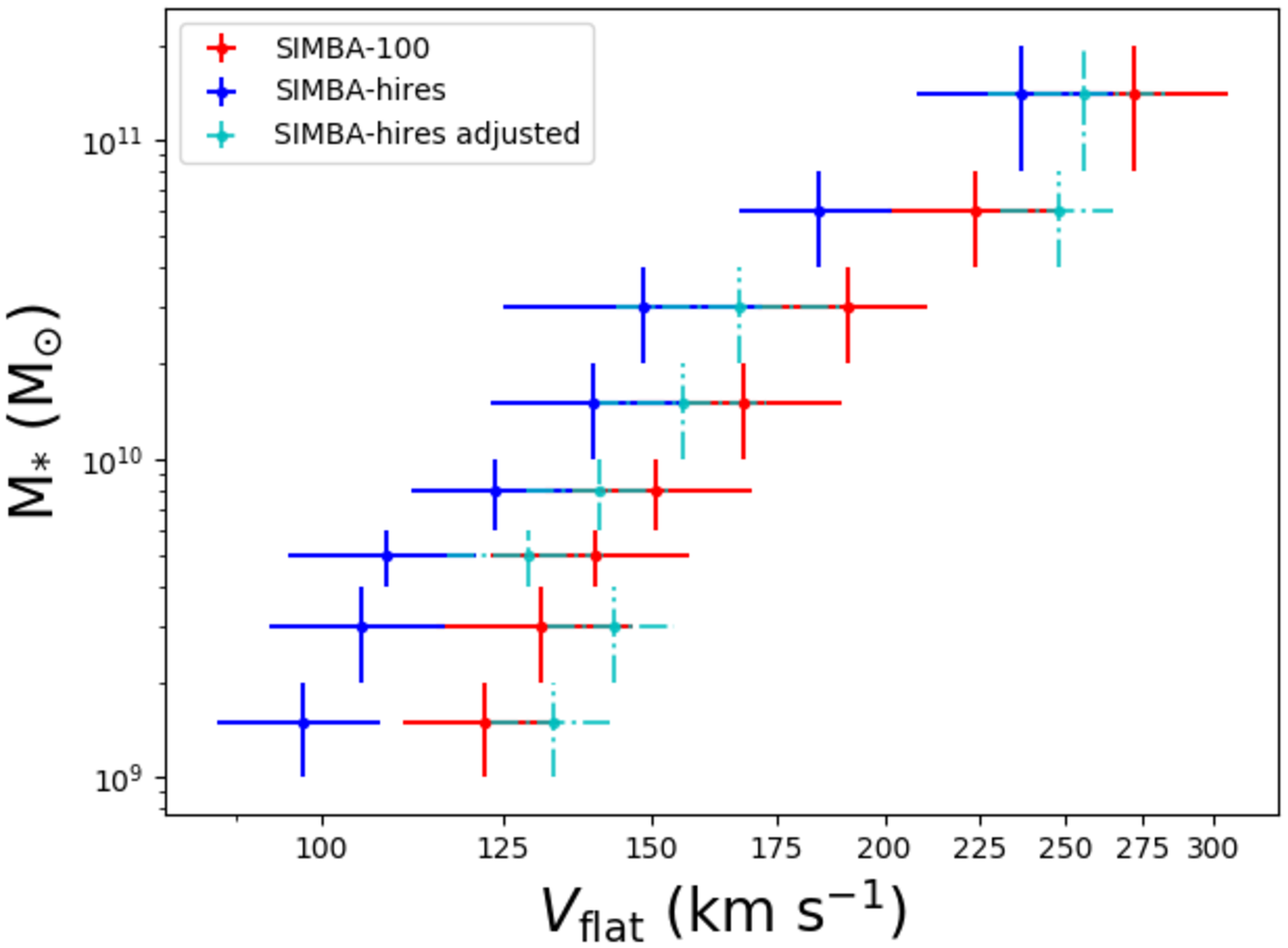}
    \caption{ Median $V_{\rm flat}$ and one-sigma errors in bins of stellar mass for the \simba\--100 and high resolution samples, in the region of velocity-space overlap between the two. We see lower rotational velocities in the high resolution snapshot. We also give the high resolution sample values when adjusting by the difference between halo and stellar masses between the two samples. Any remaining difference is attributed to numerical resolution.}
    \label{fig:mstar_velbin}
\end{figure}

We now consider similar fits for the various different definitions of rotational velocity described in Section~\ref{sec:veldefs}. We have seen that \simba\ does a reasonable job of reproducing the SPARC data for $V_{\rm flat}$, which tends to measure the velocity profile in the galaxy outskirts. Our other definitions tend to measure the velocities closer to the peak, which tend have higher values. SPARC also measured $V_{\rm max}$ and $V_{2R_e}$, although not $V_{\rm Polyex}$. In this section we compare to the BTFRs with different $V$ definitions, to understand whether \simba\ likewise reproduces these observations.

Fig.~\ref{fig:btfr} shows the $z=0$ BTFR from \simba\ for  $V_{\rm{flat}}$, $V_{\rm{max}}$, $V_{\rm{2R_{e}}}$, and $V_{\rm{polyex}}$.  In each case, the black hexbins show the individual galaxies in the full sample, with contours enclosing 0.1, 0.2, 0.5, 1, 2, 4, 6, 8, 10, and 12-$\sigma$ of the Gaussian kernel density estimation of the \simba-100 galaxy distribution. The best-fit relation for \simba\ galaxies between 100~$<$~$V$~$<$~300~km\,s$^{-1}$ is shown as the red line, in comparison with the SPARC fit as the green dashed line. Note that the SPARC fit is taken directly from \citet{Lelli2019} and is not restricted to this velocity range, although Fig.~\ref{fig:sparccompare} showed that at least for the case of $V_{\rm flat}$ the restricted fit was not significantly different.

In general, the fits from \simba\ are shifted to higher baryonic mass at a given rotational velocity, or alternatively a higher rotational velocity at a given $M_{\rm bar}$. The effect is more dramatic for the velocity measures probing the peak velocity $V_{\rm{max}}$ and is evident even in $V_{2R_e}$.  This suggests that \simba\ tends to produce too high a peak for its rotation curves, even while the outskirt velocities match reasonably well.  We note that the discrepancy, while quite evident in the plots, actually represents a fairly small deviation of $\la$0.1~dex in $V_{\rm{max}}$ or $\sim 0.03$~dex in $V_{2R_e}$ at a given $M_{\rm bar}$.  Table~\ref{tab:btfr_fits} summarises these fits by presenting the best fitting values for slope $m$, y-intercept $b$, logarithm of the y-intercept at 200~km\,s$^{-1}$ $b_{200}$, and their orthogonal intrinsic scatter $\sigma_\perp$. 

Physically, this suggests that \simba\ produces slightly overly bulge-dominated galaxies.  One reason for this may be the way feedback from star formation is implemented, via decoupled winds. By decoupling, this explicitly avoids any interaction between the wind material and ambient ISM gas. While this is intended to mimic the effects of channels of hot escaping gas from supernovae, this necessarily underestimates the effects of entrainment and energy deposition in the ISM. Decoupling also suppresses the burstiness in star formation that may be required to suppress the central dark matter density~\citep{PontzenGovernato2013}, as seen in high-resolution zoom simulations that implement the feedback more self-consistently~\citep{Brook2011,Chan2018}.  Hence simply including outflows that suppress stellar growth appropriately may not be sufficient to fully reproduce galactic structure as observed (although it gets fairly close), and it may be necessary to implement feedback in a way that explicitly impacts the ISM and central dark matter concentration in order to match all the various measures of the BTFR.  

Another possibility is that the overly concentrated galaxies are simply an effect of numerical resolution.  To check this, we compare the values from \simba-100, shown as the contours, and \simba-hires, shown as the blue points.  Indeed, we see a significant offset: in the overlapping $V$ range of $V\ga 100$~\kms, the blue points clearly lie on the high side of the contours. This shows that the BTFR is not ideally converged at the resolution of \simba-100. While there are a small number of high-$V$ galaxies in \simba-hires, their points seem to lie significantly closer to the observed BTFR relations.  However, at low $V$, the BTFR predicted in \simba-hires steepens substantially, which is not immediately evident from the observations.

Fig.~\ref{fig:btfrstats} shows the fit results from Table~\ref{tab:btfr_fits} in pictorial form, along with some other fits.  From left to right, the three panels show the results for $m$, $b_{200}$, and $\sigma_\perp$. In each case, we give results for three different data sets: the full sample (blue), galaxies with a rotational velocity between 100~$<$~$V$~$<$~300~km\,s$^{-1}$ (red), and galaxies with $V$~$>$~300~km\,s$^{-1}$ (green). Error bars show the $1\sigma$ uncertainties on the fits as calculated through the {\sc emcee} framework. We also show that the effect of adjusting the $\Upsilon_{*}$ value to match that found in SPARC does not significantly alter our best-fitting parameters; it only slightly steepens the slope.

Table~\ref{tab:btfr_fits} also gives the fits for the \simba-100 and \simba-hires individually. We note that \simba-hires gives significantly steeper slopes, which owes to the steepening of the BTFR at low-$V$ that dominates the fit. This BTFR steepening effect for low-mass galaxies was also observed in \cite{Brook2016} for MaGICC simulated galaxies, and likewise in \cite{Sales2017} from the APOSTLE/EAGLE simulations. It was proposed that a steepening in the M$_{\rm bar}$-M$_{\rm halo}$ relation at lower masses in turn drives the same steepening of the BTFR, which is an effect we also see for the higher resolution \simba\ snapshot galaxies. We note that the steepening in the BTFR for our galaxies appears to begin at a higher baryonic mass (log$_{\rm 10}$(M$_{\rm bar}$)~$\sim$~9.5\,M$_{\odot}$) than for the BTFR presented in fig. 7 of \cite{Sales2017} (around log$_{\rm 10}$(M$_{\rm bar}$)~$\sim$~8.8\,M$_{\odot}$). However, we highlight that \cite{Sales2017} probes a wider range of galaxy baryonic masses and hence goes to lower rotational velocities ($\sim$~20~km\,s$^{-1}$) than our sample ($\sim$~60~km\,s$^{-1}$), making a direct comparison on the steepening of the BTFR at the low mass end limited.

In Fig.~\ref{fig:mstar_velbin} we give the median $V_{\rm flat}$ value of galaxies in each snapshot, in bins of stellar mass. There is a lack of convergence on $V_{\rm flat}$ between the two samples, as we find lower rotational velocities for the high resolution snapshot galaxies. However, it is noted that the high resolution snapshot galaxies tend to have lower halo masses at a given stellar mass compared to the \simba\--100 snapshot, a result of keeping the same physics models for both snapshots rather than tuning for the high resolution snapshot. We take the running median of halo masses in the same stellar mass bins, and adjust the rotational velocity for the high resolution snapshot by a factor of ($M_{\rm halo-100}$/$M_{\rm halo-hires}$)$^{1/3}$, plotted in dashed cyan lines in Fig.~\ref{fig:mstar_velbin}. These give a closer agreement, with the remaining slight divergence attributed to a difference in numerical resolution.

One can identify the trends we have discussed, now presented more quantitatively. For the full or velocity-restricted samples, $V_{\rm max}$ has the steepest slope, which is not what is qualitatively seen in the SPARC sample. While this relation also has the tightest slope in \simba, the high peak which we attributed to overly large galaxy bulges makes this result in conflict with observations.  The slopes for $V_{2R_e}$ and $V_{\rm flat}$, measuring the rotation curves beyond the peak, agree significantly better.  However, the amplitudes are still too large in $V$, as a residual effect of the overly concentrated baryonic mass.

$V_{\rm max}$ also has the tightest orthogonal scatter, while \cite{Lelli2019} showed that $V_{\rm{flat}}$ actually gives the smallest scatter in the data.  \simba's $V_{\rm{flat}}$ measure gives the next tightest scatter (with the $\Upsilon_{*}$ adjustment), but it is very close to the scatter in the other measures, and is somewhat subject to assumptions regarding $\Upsilon_{*}$. This hints that perhaps rotation speeds are more tightly tied to stellar mass than to halo properties (which would reflect more in the outskirts of the rotation curve) in \simba, and suggests that while there is no large change to the best-fitting BTFR slope and intercept values found in adjusting the mass-to-light ratio, it can affect the scatter of the relation. Interestingly, despite using the full range of $\Upsilon_{*}$ values directly from \simba, the scatter is actually reduced relative to using fixed values of $\Upsilon_{*}$.  This is supported by observational studies that indicate that rotation speeds are more tightly tied to stellar mass than to halo properties \cite[e.g.][]{McGaugh2012}.  We will examine the connection of circular velocities to halos in future work.

An interesting sidelight is the best-fit values for the $V>300$\kms\, samples, which are akin to the super-spirals of \cite{Ogle2019}. These are systematically deviant from the bulk of the sample, showing a substantially lower slope in almost all measures, which is qualitatively in agreement with observations.  However, the small number of such galaxies in \simba\ results in quite large formal uncertainties, so that it is typically not discrepant by more than $\sim 1\sigma$. In addition to highlighting the importance of comparing to velocity-restricted samples, it is interesting to understand physically why such objects deviate from the standard relation.  One possibility is that the baryonic mass in these systems as measured via neutral gas is not increasing at the same rate with halo mass as seen in lower-mass systems, because such large galaxies tend to have more of their gas in hot form that would not be visible in \HI.  Hence perhaps if one could observe the baryonic mass from all the gas rather than just the \HI, such as including the hot gas from X-ray measurements~\citep{Anderson2015}, this would raise $M_{\rm bar}$ in these systems and alleviate this discrepancy.

Overall, we see that \simba\ does a reasonable job of reproducing SPARC observations of the BTFR in galactic outskirts as measured by $V_{\rm flat}$, but it tends to produce too large peak rotation speeds at a given baryonic mass.  The discrepancy is fairly modest, $\la 0.1$~dex in velocity, but at face value it suggests that \simba's feedback mechanism does not suppress central bulge growth quite as much as observed. It is beyond the scope of this work to examine a bulge/disk decomposition for these simulated galaxies to test this hypothesis more directly, but if this is true, then it may indicate that simply removing gas via outflows is not entirely sufficient to reproduce the BTFR, but feedback must also energetically impact the central regions of galaxies. Additionally, it appears that \simba\ rotation velocities are not well converged at the resolution of the main $100\hmpc$ volume, with the higher-resolution volume showing offsets to higher $V$ comparable to the $0.1$~dex discrepancy, albeit with a steeper slope.  Nonetheless, the broad success of \simba\ in matching the BTFR in its various forms is encouraging, and is certainly a significant improvement over the earliest simulations that did not include outflows.

\subsection{Removing bulge-dominated galaxies}\label{sec:nobulge}

It should be noted that the SPARC survey included galaxies that are disk-dominated, with the sample collected from previous studies. The \simba\ samples presented here thus far have not been chosen to be specifically disk-dominated; rather, we have merely required a minimum in the H{\sc i} and stellar content, and sufficient sSFR - attributes that are true of many spiral galaxies, but are not exclusive to them. As such, \simba\ galaxies with large bulges that are not present in SPARC may affect our results, such as contributing to the observed systematic shift in velocities with respect to SPARC. 

While a detailed comparison of morphologies is beyond the scope of this paper, we conduct a simplistic morphology estimator for our \simba\ galaxies, to investigate the BTFR parameters when removing galaxies with a bulge `contamination'. Two measures computed in \caesar\ are considered for the galaxy gas particles, in order to compare directly to SPARC's rotation curve values derived solely from H{\sc i} content:

\begin{itemize}
    \item Bulge mass over total mass (B/T) - where the bulge mass of the galaxy is, assuming a non-rotating bulge, twice the mass of particles counter-rotating with respect to the momentum vector of the galaxy. 
    \item $\kappa_{\rm rot}$  - the fraction of kinetic energy invested in ordered rotation, as used in \cite{Sales2012} for Millennium Simulation galaxies. In that work, they characterise galaxies with $\kappa_{\rm rot}$~$<$~0.5 as spheriod-dominated. 
\end{itemize}

In the second page of Table~\ref{tab:btfr_fits}, we give the best fitting BTFR parameters for \simba\ galaxies with a limit of B/T~$<$~0.3 imposed, and separately $\kappa_{\rm rot}$~$>$~0.5, to exclude galaxies dominated by their bulge content. We find for both cases there is only a small difference in the slope and intercept across all velocity definitions compared to when no morphological cut is made, and thus that the presence of bulge-dominated galaxies is not skewing our results significantly. Similar values are found when considering stellar particles in the B/T and $\kappa_{\rm rot}$ calculation for each galaxy, as well as both stellar and gas particles together. Furthermore, similar BTFR parameters are found under different velocity limits on the full sample, and for the two individual snapshot samples separately. We note that the best-fitting parameters will vary slightly based on where the morphology estimation cuts are made.

\subsection{Mass dependence of rotation curves}\label{sec:massdep}

\begin{figure}
    \centering
  \includegraphics[width=0.46\textwidth]{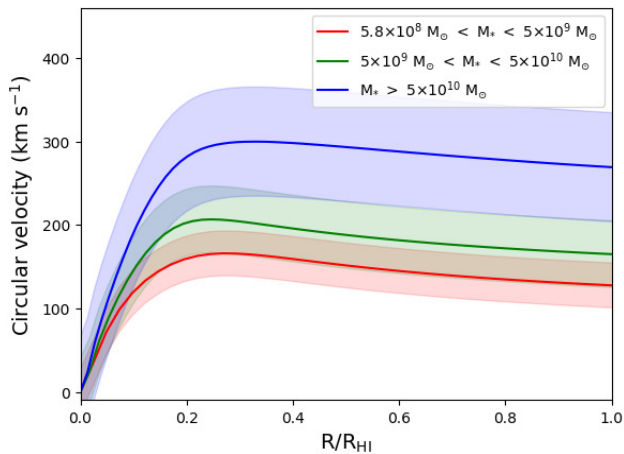}
    \caption{Binned rotation curves for the 100~Mpc\,h$^{-1}$ snapshot, with high stellar mass galaxies (M$_{*}$~$>$~5\,$\times$\,10$^{10}$\,M$_{\odot}$) in blue, intermediate stellar mass galaxies (5\,$\times$\,10$^{9}$\,M$_{\odot}$~$<$~M$_{*}$~$<$~5\,$\times$\,10$^{10}$\,M$_{\odot}$) in green, and low stellar mass galaxies (5.8\,$\times$\,10$^{8}$\,M$_{\odot}$~$<$~M$_{*}$~$<$~5\,$\times$\,10$^{9}$\,M$_{\odot}$) in red. Error bounds give the 1-$\sigma$ range for each bin. The extent of the binned rotation curves is to the median distance determined by the \HI\ mass-size relation for each bin. Larger mass \simba\ galaxies have higher circular velocity peaks, unsurprisingly, but also have flatter rotation curves, compared to the lower mass galaxies, whose peaks are more pronounced and have a sharper decline in their rotation curve (see Table~\ref{tab:velmasscompare}).}
    \label{fig:massbinrotcurves}
\end{figure}

We expect that the larger stellar-mass galaxies also have a larger total baryonic mass, and thus a higher rotational velocity according to the BTFR.  A more interesting question is, is there a systematic difference in the shapes of the rotation curves as a function of mass?  We investigate this question by separating our sample into three stellar mass bins between 5.8\,$\times$\,10$^{8}$\,M$_{\odot}$, 5.8\,$\times$\,10$^{9}$\,M$_{\odot}$ and 5\,$\times$\,10$^{10}$\,M$_{\odot}$, and constructing an average rotation curve for each bin.  We use stellar mass bins because this provides an additional piece of information on the BTFR, since it also depends on the gas-to-stellar ratio.

Fig.~\ref{fig:massbinrotcurves} presents the results of the median rotation curve in these mass bins.  The solid lines show the median values for high, intermediate, and low stellar mass bins (blue, green, and red lines) as indicated, and the shaded region encompasses the $\pm1\sigma$ range of circular velocity values at the given radius. The rotation curves extend to the median radius determined by the \HI\ mass-size relation for each bin; this distance is greater for the higher mass galaxies, as they have a larger \HI\ extent.

High stellar mass galaxies have higher circular velocities than lower stellar mass galaxies, as expected.  What is more surprising is that, up to the radius cut-off we take for each \simba\ galaxy, larger mass galaxies have flatter rotation curves, while smaller mass galaxies have a more pronounced inner peak.  This is generally contrary to observations that show that dwarf galaxies often have dark matter cores that result in slowly rising rotation curves, compared to Milky Way-sized galaxies~\citep[e.g.][]{Swaters2002}. 

To quantify this, Table~\ref{tab:velmasscompare} gives the median velocity at the end of the rotation curves, the median value at the peak ($V_{\rm peak}$), and the ratio between the two, for each mass bin.  It can be seen that in the highest mass bin, the rotation curve drops by about 11\% from the peak to the outskirts, while for the lowest mass bin this drop is 24\%.

\begin{table}
    \centering
    \begin{tabular}{lccc}
    \hline 
Stellar mass & $V_{\rm{max}}$ & $V_{\rm{end}}$ & $\frac{V_{\rm{end}}}{V_{\rm{max}}}$ \\
M$_{\odot}$ & km\,s$^{-1}$ & km\,s$^{-1}$ & \\
\hline 
5.8\,$\times$\,10$^{8}$~$<$~M$_{*}$~$<$~5\,$\times$\,10$^{9}$ & 165.9 & 125.5 & 75.6\% \\
5\,$\times$\,10$^{9}$~$<$~M$_{*}$~$<$~5\,$\times$\,10$^{10}$ & 206.4 & 162.4 & 78.6\% \\
5\,$\times$\,10$^{10}$~$<$~M$_{*}$ & 299.8 & 265.7 & 88.6\% \\
\hline
\end{tabular}
    \caption{Comparison of the peak velocity, velocity at the end of the rotation curve (defined through the \HI\ mass-size relation), and the ratio $V_{\rm{end}}$:$V_{\rm{max}}$, for the average rotation curves for galaxies in stellar mass bins (Fig.~\ref{fig:massbinrotcurves}). Lower stellar-mass galaxies have a greater decrease in velocity from the peak compared to high mass galaxies.}
    \label{tab:velmasscompare}
\end{table}

It appears clear that \simba, while reproducing the overall BTFR reasonably well for $V_{\rm flat}$, does not yield the observed mass dependence in the shape of the galaxy rotation curve. In particular, it suggests that in the inner regions of low-mass galaxies, the dark matter contribution is higher than expected.  Indeed, this corroborates the trend we saw earlier for the different velocity measures, where \simba\ produced too high values of $V_{\rm peak}$ at a given baryonic mass. The underlying physical reason(s) for this discrepancy are as outlined previously.  

\begin{figure*}
\begin{minipage}{\textwidth}
\centering
\begin{subfigure}[b]{0.475\textwidth}
  \includegraphics[width=\textwidth]{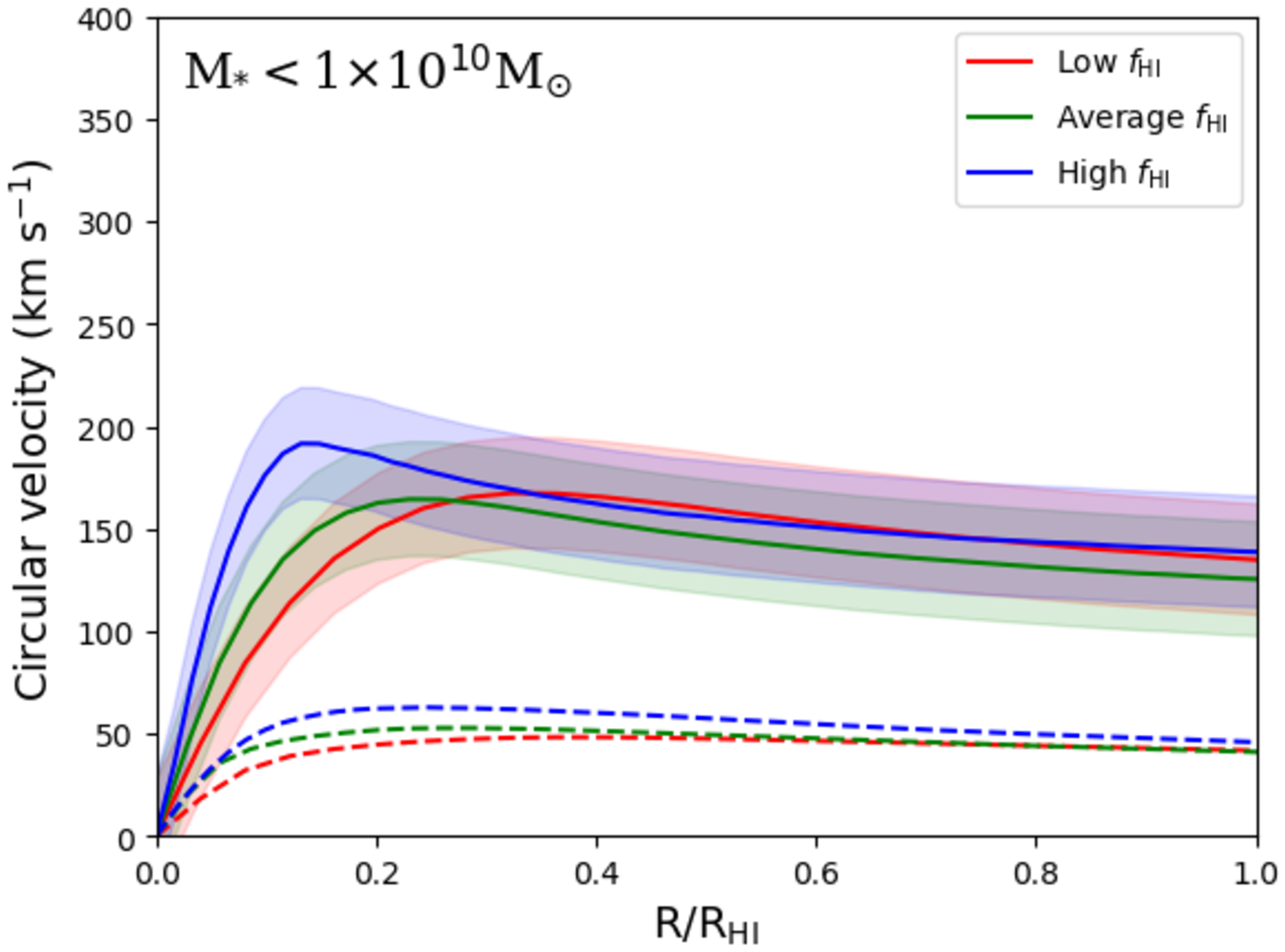}
\end{subfigure}
\begin{subfigure}[b]{0.475\textwidth}
  \includegraphics[width=\textwidth]{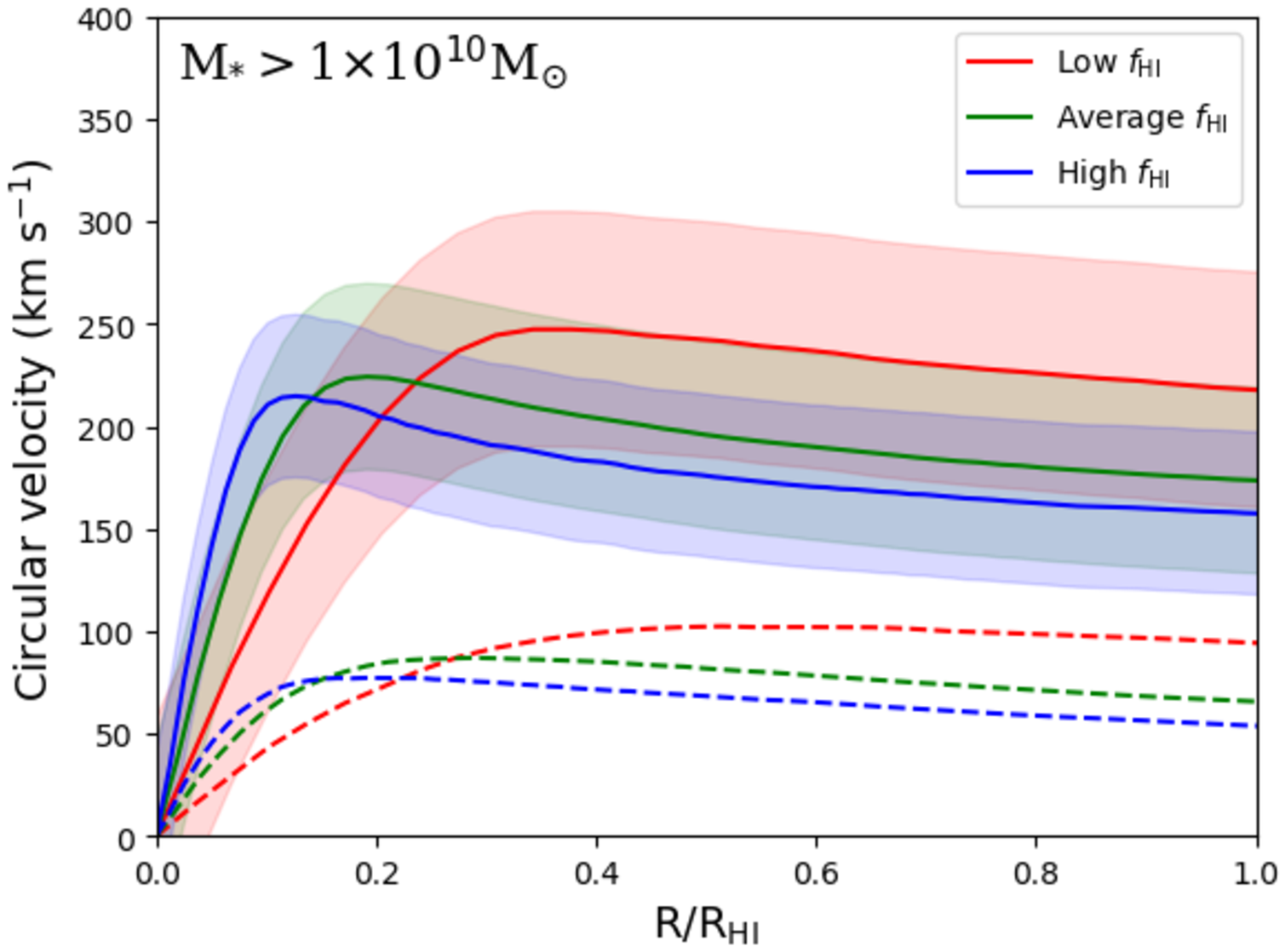}
\end{subfigure}
\end{minipage}
\caption{Binned rotation curves by \HI\ gas fraction (\HI\ mass divided by stellar mass). As in Fig.~\ref{fig:massbinrotcurves}, the rotation curves extend to the median \HI\ mass-size radius for each bin, and otherwise peak at roughly the same radius in kpc. The left panel includes all galaxies with stellar masses below 1\,$\times$\,10$^{10}$\,M$_{\odot}$, while the right contains higher-mass galaxies (above 1\,$\times$\,10$^{10}$\,M$_{\odot}$). For these high-mass} galaxies, we see higher rotational velocities with lower \HI\ gas fractions, but this trend disappears when including less massive galaxies. The dashed lines give the contribution of the baryonic material to the rotation curves.
\label{fig:himassbinrotcurves}
\end{figure*}




\begin{figure*}
    \centering
    \includegraphics[width=1.\linewidth]{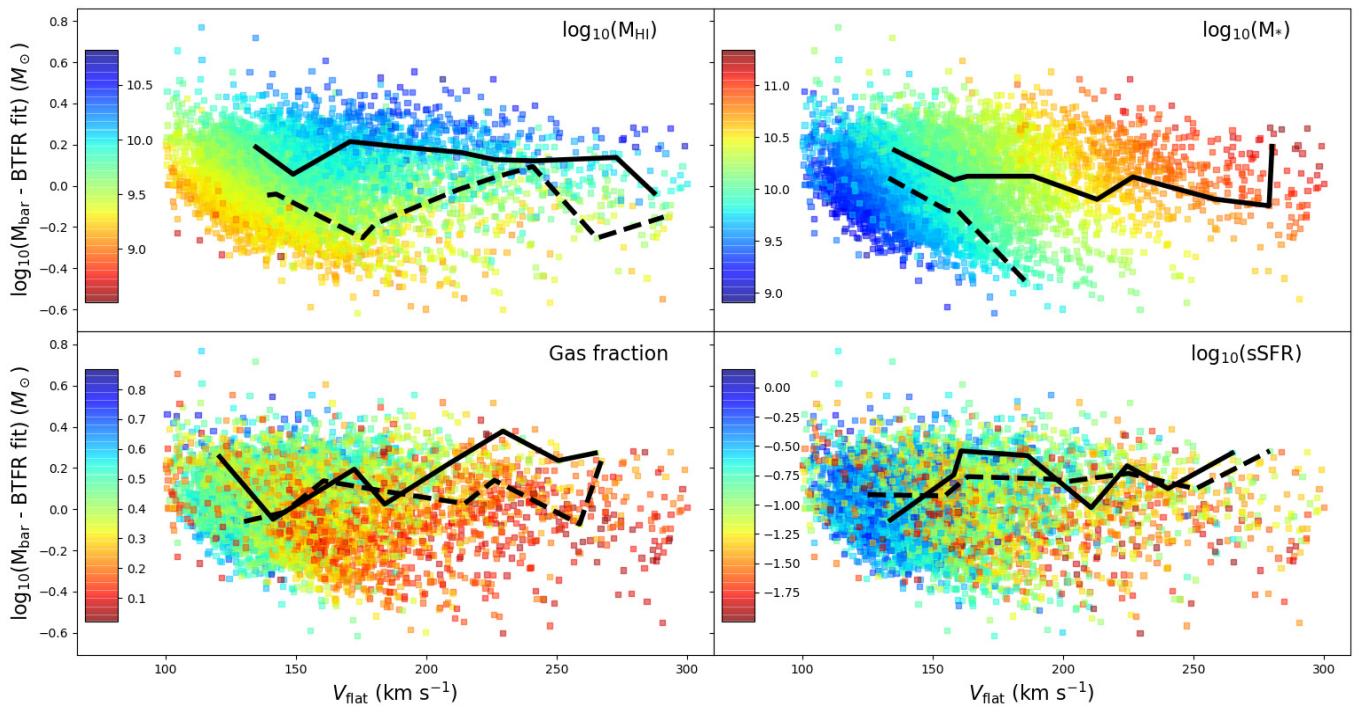}
    \caption{Comparison of the distribution of the HI mass, stellar mass, gas fraction, and sSFR across the BTFR for the $V_{\rm{flat}}$ definition. Here we have subtracted the best fit from the BTFR we obtained for velocities between 100 and 300~km\,s$^{-1}$. We give the running median for points above and below the best fit (i.e. values of 0 on the y-axis) in solid and dashed black lines respectively.}
    \label{fig:btfrminfit}
\end{figure*}

\subsection{HI fraction dependence of rotation curves}\label{sec:hifractiondep}

One concern in conducting the comparison between \simba\ and observations such as SPARC is that the galaxy selection function is different. In \simba, the selection is primarily based on cuts in stellar and \HI\ mass, since this provides the most robust selection against potential numerical resolution issues. In observations, the selection is typically more heterogeneous, but particularly for samples that probe rotation curves to large radii using \HI, it requires that galaxies be fairly \HI-rich in order to be able to make such measurements. It is thus a potential concern that galaxies that have large $\fhi=\mhi/M_*$ are somehow systematically biased in the shape and/or amplitude of their rotation curves, and this could lead to apparent discrepancies such as the one regarding the shapes of the rotation curves as discussed in the previous section.  

In this section, we examine the dependence of rotation curves on whether galaxies have high or low \HI\ fractions for their given stellar mass. To do this, we first compute a running median relation between $\fhi$ and $M_*$, the `\HI\ main sequence'. Then, we construct three binned populations in percentiles relative to the \HI\ main sequence: those that are $<25$\% below, $>75$\% above, and in between. Finally, we then determine the average rotation curves of those populations.

Fig.~\ref{fig:himassbinrotcurves} shows the results of this.  It turns out that the results are qualitatively somewhat different depending on the mass of the galaxy, so in the left panel we consider our full sample with M$_{*}<10^{10}M_\odot$, while in the right panel we limit to more massive galaxies (M$_{*}>10^{10}M_\odot$). In each panel, the median rotation curve for the low, intermediate, and high $\fhi$ galaxies are indicated in red, green, and blue, respectively. The shadings show the $1\sigma$ range around the median. The coloured dashed lines give the baryonic matter contribution to the median rotation curves for each subsample.

In the left panel, we do not see a strong trend in rotation curve amplitude for the low-mass galaxies in the outer parts of the rotation curves. There is a modest tendency for the highest and lowest $\fhi$ systems to have higher circular velocities by $\sim~20$\kms~in the outer parts, relative to galaxies with average $\fhi$ values. Thus samples that select dwarf galaxies having high $\fhi$ but relatively low stellar masses will tend to bias towards larger circular velocities than the typical galaxy at that mass. This is a weak effect but could be important for precise comparisons.

We note that there is a higher peak at inner radii for low-mass galaxies with high $\fhi$ values. However, we highlight that the x-axis is scaled by the median radius extent, determined by the H{\sc i} mass-size relation; low and average $\fhi$ systems have a median half-mass radius of 3.7~$\pm$~0.8 and 3.5~$\pm$~0.9~kpc respectively, while high $\fhi$ systems have half-mass radii of 5.0~$\pm$~0.9~kpc. The general shape of the baryonic contribution to these rotation curves (dashed lines) do not differ much between the three subsamples, suggesting that it is the dark matter distribution that dominates here for dwarf galaxies.

In the high-mass galaxies (right panel), we see an interesting effect that there is now a strong trend: galaxies with the lowest $\fhi$ now have the highest $V$, by $\sim 40$\kms above the typical galaxy beyond the peak. We note that for the low $\fhi$ systems, we find a slightly higher median dark matter halo mass of 1.4~$\times$~10$^{12}$~M$_\odot$, compared to 9.7~$\times$~10$^{11}$~M$_\odot$ and 9.3~$\times$~10$^{12}$~M$_\odot$ for the average and high $\fhi$ systems respectively. When considering galaxies with H{\sc i} masses above 5~$>10^{9}M_\odot$ instead of a stellar mass cut, to investigate if there is a difference for galaxies with large neutral gas quantities, the difference is even more pronounced: low $\fhi$ systems have a median dark matter halo mass of 2.3~$\times$~10$^{12}$~M$_\odot$, typical $\fhi$ systems have 8.2~$\times$~10$^{11}$~M$_\odot$, and high $\fhi$ systems 7.0~$\times$~10$^{12}$~M$_\odot$. The same trend in rotation curves is also observed in this cut. Hence, the effect we see here may be driven by the difference in dark halo mass; that is, galaxies with large dark halo masses have both higher rotation curves and lower values of $\fhi$. This may in turn be linked to higher satellite abundances found for isolated galaxies with more massive dark matter haloes \citep{Wang2012}. This is an important prediction that can be further tested in upcoming H{\sc i} observational studies

What is not evident is any difference in the general shapes of the rotation curves as a function of position relative to the \HI\ main sequence. This rules out \HI\ selection bias as an explanation for why \simba\ dwarf galaxies tend to have peakier rotation curves than observed (Section~\ref{sec:massdep}), and suggests instead that this discrepancy is a true failing of the model.

As an aside, we also examined differences in the rotation curves versus position with respect to the star-forming main sequence, i.e. using sSFR to subdivide the samples rather than $\fhi$.  There was no trend seen -- galaxies with high, intermediate, and low sSFR had essentially exactly the same median rotation curve and spread around it. We do not show this plot since the result is easily summarized. This indicates that while galaxy rotation curves have some second-parameter dependence on \HI\ fraction when controlling for stellar mass, they have none versus sSFR.

The second parameter dependence on $\fhi$ seen in Fig.~\ref{fig:himassbinrotcurves} begs the question as to why this occurs, and in particular why the sense of the trend is different for low-mass versus higher-mass galaxies. To better understand this, we investigate whether there are additional trends that could also help explain this result, and if any dependencies exist in the residual of the BTFR.


\subsection{BTFR second parameter dependencies}\label{sec:btfrcategories}

To further break down the dependences of the BTFR on galaxy properties, we now examine these properties as a function of distance from the BTFR fit.  In other words, do galaxies that lie above the BTFR at a given baryonic mass have systematically different properties than those that lie below? And do any trends exist that have a dependence on the circular velocity?

Fig.~\ref{fig:btfrminfit} shows the results of such an analysis.  Here we show a scatter plot of the deviation in $M_{\rm bar}$ from the best-fit BTFR versus the circular velocity $V_{\rm{flat}}$, colour-coded by four different quantities: $\mhi$ (upper left), $M_*$ (upper right), gas fraction (lower left), and sSFR (lower right).  We show two median lines: the solid black line shows the running median versus circular velocity of all galaxies that are in the upper half of each given quantity (i.e. the blue/green part for the $\mhi$ case), while the dashed black line shows the running median in the lower half (e.g. the yellow/red points for $\mhi$); these are effectively the running 75\% and 25\% percentile lines, respectively. Therefore, from bottom to top, we can see the impact of increasing baryonic mass at a given rotational velocity, while from left to right we can see if these trends are correlated with rotational velocity.

Looking at $\mhi$, we see a clear vertical trend that galaxies that lie above the BTFR (i.e. above 0 on the $y$-axis) also tend to have high $\mhi$ (bluer). This is not too surprising, since the baryonic mass includes $\mhi$.  For most velocities, the upper half of $\mhi$ galaxies has $\sim0.1-0.3$~dex higher values of $M_{\rm bar}$ than the lower half. 

The $M_*$ panel (upper right) broadly shows the unsurprising trend that higher $M_*$ galaxies have higher circular velocity. But even at a given velocity, there is a trend that galaxies lying above the BTFR tend to have higher $M_*$.  Again, since $M_{\rm bar}$ includes $M_*$, this is qualitatively expected. So e.g. at 200~\kms, galaxies range from $M_*\sim 5\times 10^{10}M_\odot$ at the highest $M_{\rm bar}$, down to $\sim 10^{10}M_\odot$ at the lowest $M_{\rm bar}$.  Hence a galaxy's location in the BTFR plane is dependent on both its $\mhi$ and $M_*$.

The lower panels show less obviously discernible trends. Looking at the gas fraction panel in the lower left, only at higher velocities do we see that galaxies with low gas fraction tend to lie below the BTFR at velocities above 200~\kms\,(i.e. the dashed line lies below the solid line). A likely explanation for this is that for larger galaxies, the galaxies are stellar mass dominated, and larger stellar mass galaxies tend to have lower gas fractions since they are more often quenched~\citep{Dave2019}. Indeed, these galaxies are large enough to be quenched and yet remain above the BTFR best-fit. This is an interesting, albeit weak, prediction for how quenching is manifest in the residuals of the BTFR.


From this trend of low gas fraction at high rotational velocities, we can go back and interpret Fig.~\ref{fig:himassbinrotcurves} where we found that for galaxies with stellar masses greater than 1\,$\times$\,10$^{10}$\,M$_{\odot}$ the galaxies with lower \HI\ gas fractions have higher rotational velocities, but galaxies with smaller stellar masses have similar rotation curves to those with higher \HI\ gas fractions. The division of $M_{*}~=~1\times\,10^{10}\,M_{\odot}$ approximately corresponds to 180--200~km\,s$^{-1}$ (top-right panel). Hence, when considered together with the bottom-left panel, low $M_{*}$ galaxies with high gas fraction do not differ substantially from low M$_{*}$ galaxies with low gas fraction, while at higher M$_{*}$ masses, low gas fraction galaxies dominate as high gas fraction sources drop out beyond $V$~$\sim$~200~\kms. We note the same trend in Fig.~\ref{fig:himassbinrotcurves} exists when making a cut in H{\sc i} mass (Section~\ref{sec:hifractiondep}); the most H{\sc i} massive galaxies (top-left panel) appear at higher velocities as well, albeit less prominently compared to stellar mass dependence.

For the sSFR (lower right), there is the general trend that galaxies with higher sSFR tend to have small velocities, i.e. they are smaller systems; again this is unsurprising. We do not detect a significant residual trend besides this.  In general, we have found that the BTFR is quite insensitive to sSFR.  In part, this probably owes to our selection which isolates star-forming galaxies with significant neutral gas. 

Overall, we see that galaxies show expected trends in \HI\ and stellar mass in the BTFR residuals given their contribution to the baryonic mass, and only find a slight trend for low gas fraction galaxies to dominate at rotational velocities above 200~km\,s$^{-1}$. No additional trends were apparent when applying the simple morphological indicator cuts discussed in Section~\ref{sec:nobulge}. The inversion in the gas fraction dependence relative to $\fhi$ is particularly interesting as it corresponds to the onset of quenching when \HI\ begins to be removed from galaxies and they become more stellar dominated. Examining these trends versus future observations will provide more detailed constraints on galaxy evolution models.



\section{Summary}\label{sec:summary}

We have examined the Baryonic Tully Fisher Relation (BTFR) in the \simba\ cosmological galaxy formation simulation at $z=0$, and compared where possible to the SPARC survey of \citet{Lelli2016}.  To do so, we generated rotation curves for a sample of gas-rich \simba\ galaxies, and analysed these by constructing `ideal' rotation curves from the mass distribution, which we confirmed closely followed the rotation curve obtained from the gas in these rotationally-dominated systems. We examined the BTFR for four different definitions of circular velocity that have appeared in the literature: $V_{\rm{flat}}$, $V_{\rm{max}}$, $V_{\rm{2R_{e}}}$, and $V_{\rm{polyex}}$.  Furthermore, we have investigated second parameter trends for the BTFR with respect to key global galaxy properties.

Our main results follow.

\begin{itemize}
    \item Using $V_{\rm{flat}}$ as was favoured by the SPARC survey \citep{Lelli2016}, we find good agreement in the BTFR relation between \simba\ and SPARC. In detail the \simba\ relation is slightly shallower than observed, with a $V_{\rm{max}}$ slope of $3.5$ rather than $\approx 3.8$, and an amplitude offset towards higher $V_{\rm{flat}}$ of $\la 0.07$~dex.
    
    \item For $V_{\rm{max}}$ and $V_{\rm{2R_{e}}}$, \simba\ tends to produce steeper BTFR slopes than observed, and predicts higher $V$ at a given baryonic mass than seen in SPARC, though again the discrepancy is modest ($\la 0.1$~dex in $V$).  This suggests that \simba\ over-predicts the velocities closer to the peak of the rotation curve, while the velocities in the outskirts ($V_{\rm{flat}}$) are in better agreement with data.
    
    \item We test the Polyex model for fitting rotation curves, and find that it is a viable method that provides similar BTFR results to $V_{\rm{max}}$.

    \item \simba\ qualitatively reproduces the trend of a flatter BTFR for $V$~$>$~300~km\,s$^{-1}$, as seen in super spiral galaxies~\citep{Ogle2019}, though the statistics are too low to provide strong constraints.
    
    \item  For the orthogonal scatter around the BTFR, we find the tightest and steepest BTFR using $V_{\rm{max}}$, while $V_{\rm{flat}}$ is somewhat less tight.  The scatter values are comparable to that seen for the SPARC data ($\sim 0.06-0.07$~dex), but they found that $V_{\rm{flat}}$ was the tightest.
    
    \item In \simba, galaxies with high stellar masses have slightly less peaked rotation curves than galaxies with lower stellar masses.  This trend is broadly contrary to observations.  
    
    \item A potential cause of this discrepancy, as well as the discrepancies in the $V$ values, is that \simba\ has somewhat overly concentrated mass distributions compared to real galaxies, especially in low-$V$ galaxies.
    
    \item A possible physical explanation is that feedback in \simba\ explicitly does not impact the ambient ISM upon ejection; simulations where this is accounted for tend to reproduce the radial acceleration relation~\citep[RAR;][]{Lelli2017} better~\cite[e.g.][]{Ludlow2017,Tenneti2018}.  Hence while \simba's kinetic feedback model provides a very good match to a wide range of global galaxy properties, the internal redistribution of baryons associated with such feedback may not be properly captured.
    
    \item Another possible issue is numerical resolution.  The $8\times$ higher mass resolution \simba-hires run shows higher $V$ values at a given baryonic mass than the main \simba-100 run, in a sense that tends to improve agreement with observations for more massive galaxies.  However, it also predicts a quite steep slope for the low-$V$ BTFR that is not seen in data.  In general, \simba's main $100\hmpc$ volume is not ideally resolution converged for BTFR studies, though it is not grossly impacted by this.

    
    \item  The residuals above and below the BTFR correlate with expected galaxy properties in \simba.  Galaxies above the BTFR tend to have higher $\mhi$ and higher $M_*$.  However, there is little trend to be seen with gas fraction or sSFR in this regard.
    
    \item An interesting trend is found when splitting rotation curves by \HI\ gas fractions for galaxies: high stellar-mass galaxies ($M_{\rm{*}}$~$>$~1\,$\times$\,10$^{10}$\,M$_{\odot}$) with low \HI\ gas fractions have higher rotational velocities, while the trend is suppressed when including galaxies with lower \HI\ masses.  
    
    \item This is corroborated by examining the residual deviation from the BTFR versus gas fraction, which shows that below 200~\kms, galaxies with high $\fhi$ lie above the BTFR, whereas above 200~\kms, galaxies with high $\fhi$ lie below the BTFR.  The small difference arises owing to the impact of quenching in massive galaxies, which causes these galaxies to have low $\fhi$ although they have high baryonic (mostly stellar) mass. These trends provide testable predictions using forthcoming \HI\ Tully-Fisher surveys.
    
\end{itemize}

This work broadly demonstrates that the \simba\ cosmological simulation suite is a useful tool for investigating the BTFR in sizeable galaxy samples, despite requiring careful consideration of resolution convergence issues.  Higher resolution cosmological or zoom simulations will be useful for more rigorously assessing the impact of resolution.  Moreover, it will enable investigation of more observables such as the radial acceleration relation~\citep{Lelli2017} and other measures quantifying the internal dynamics and structures of disk galaxies.  Nonetheless, the overall success of \simba\ in reproducing the BTFR to within $\sim 0.1$~dex suggests that it is a valuable platform for investigating the nature of the BTFR and its connection to halos, which we will do in future work.

\simba, alongside future \HI\ surveys such as LADUMA and MIGHTEE, will better characterise the BTFR, including obtaining larger samples and quantifying its evolution out to $z\sim 1+$.  Future work with \simba\ will aim to investigate the redshift evolution of the BTFR, while quantifying the biases in utilising spectral emission line widths to measure the BTFR. We will also explore the radial acceleration relation as an independent constraint on our simulations. These studies will set the stage for using \simba\ to explore the evolution of disk dynamical properties and their connections with halos over cosmic time in forthcoming \HI\ surveys.

\section*{Acknowledgements}
We thank the anonymous referee for comments that have helped improve the paper. We also thank Federico Lelli, Alice Concas, Andrew Baker, and Kyle Oman for helpful conversations.
We thank Robert Thompson for developing {\sc Caesar}, and the {\sc yt} team for development and support of {\sc yt}.
RD acknowledges support from the Wolfson Research Merit Award program of the U.K. Royal Society.
Ed Elson acknowledges that this research is supported by the South African Radio Astronomy Observatory, which is a facility of the National Research Foundation, an agency of the Department of Science and Technology. Marcin Glowacki acknowledges support from the Inter-University Institute for Data Intensive Astronomy (IDIA). The computing equipment to run \simba\ was funded by BEIS capital funding via STFC capital grants ST/P002293/1, ST/R002371/1 and ST/S002502/1, Durham University and STFC operations grant ST/R000832/1. DiRAC is part of the National e-Infrastructure.  We acknowledge the use of computing facilities including the visualisation lab of IDIA for part of this work. IDIA is a partnership of the Universities of Cape Town, of the Western Cape and of Pretoria.

\subsection{Data availability}

The data underlying this article will be shared on reasonable request to the corresponding author. \simba\ snapshots and galaxy catalogs used for this work can be found at \url{http://simba.roe.ac.uk/}.





\footnotesize{
  \bibliographystyle{mnras}
  \bibliography{bibliography}
}




\bsp	
\label{lastpage}
\end{document}